\documentclass[12pt,preprint]{aastex}
\usepackage{times,mathptm,graphicx,epsfig} 

\slugcomment{\today } 
\shorttitle{Chandra Multiwavelength Project} \shortauthors{Kim et al.}
\begin{document}

\title{Chandra Multiwavelength Project \\
X-ray Point Source Catalog}

\author{Minsun Kim\altaffilmark{1,2}, 
Dong-Woo Kim\altaffilmark{1},
Belinda J. Wilkes\altaffilmark{1}, 
Paul J. Green\altaffilmark{1}, 
Eunhyeuk Kim\altaffilmark{1}, 
Craig S. Anderson\altaffilmark{1},
Wayne A. Barkhouse\altaffilmark{3},
Nancy R. Evans\altaffilmark{1}, 
$\rm{\check{Z}}$eljko Ivezi$\rm{\acute{c}}$\altaffilmark{4},
Margarita Karovska\altaffilmark{1}, 
Vinay L. Kashyap\altaffilmark{1},
Myung Gyoon Lee\altaffilmark{2},
Peter Maksym\altaffilmark{5},
Amy E. Mossman\altaffilmark{1},
John D. Silverman\altaffilmark{6},
and Harvey D. Tananbaum\altaffilmark{1}
}
\email{mkim@cfa.harvard.edu}
\altaffiltext{1}{Harvard-Smithsonian Center for Astrophysics, 
60 Garden Street, Cambridge, MA 02138, USA}
\altaffiltext{2}{Department of Physics and Astronomy, Astronomy Program, 
Seoul National University,
Seoul 151-742, Korea}
\altaffiltext{3}{Department of Astronomy, University of Illinois at Urbana-Champaign, 
Urbana, IL 61801, USA}
\altaffiltext{4}{Department of Astronomy, University of Washington, Seattle, WA 98195, USA}
\altaffiltext{5}{Department of Physics and Astronomy, Northwestern University,
Evanston, IL 60208, USA}
\altaffiltext{6}{Max-Planck-Institut f\"ur extraterrestrische Physik, D-84571
Garching, Germany}

\begin{abstract}
We present the $Chandra$ Multiwavelength Project (ChaMP) X-ray point
source catalog with $\sim 6,800$ X-ray sources detected in 149
$Chandra$ observations covering $\sim 10~deg^{2}$.  The full ChaMP
catalog sample is seven times larger than the initial published 
ChaMP catalog. The
exposure time of the fields in our sample ranges from $0.9$ to $124$ ksec,
corresponding to a deepest X-ray flux limit of $f_{0.5-8.0} = 9\times
10^{-16}~erg~cm^{-2}~sec^{-1}$. The ChaMP X-ray data have been uniformly reduced and
analyzed with ChaMP-specific pipelines, and then carefully validated
by visual inspection. The ChaMP catalog includes X-ray photometric
data in 8 different energy bands as well as X-ray spectral
hardness ratios and colors.

To best utilize the ChaMP catalog, we also present the source
reliability, detection probability and positional uncertainty.  To
quantitatively assess those parameters, we performed extensive
simulations. In particular, we present a set of empirical equations: 
the flux limit as a function of
effective exposure time, and the positional uncertainty as a function
of source counts and off axis angle. The false source detection rate
is $\sim 1 \%$ of all detected ChaMP sources, while the detection
probability is better than $\sim95\%$ for sources with
counts $\gtrsim 30$ and off axis angle $<5\arcmin$. 
The typical positional offset between ChaMP
X-ray source and their $SDSS$ optical counterparts is $0.7\pm0.4\arcsec$,
derived from $\sim900$ matched sources. 
\end{abstract}

\section{Introduction}
Since the cosmic X-ray background (hereafter CXRB) was discovered by \citet{gia62},
there have been several X-ray missions such as $Einstein$, $ROSAT$,
$ASCA$ and $BeppoSAX$.
The $Chandra$ $X$-$ray$ $Observatory$
(hereafter $Chandra$) and the $X$-$ray$ $Multi$-$Mirror$ $Mission$-$Newton$ 
(hereafter $XMM$-$Newton$) are the current powerful X-ray missions with
more sensitive imaging spectroscopy and higher positional accuracy 
than previous missions.
To investigate the formation and evolution of galaxies, clusters of
galaxies, and large scale structure of the universe, these previous and
current X-ray missions have provided  
several deep and wide extra galactic X-ray surveys 
(see \citet{bra05} and references therein for a detailed review).
\citet{bau04} established the X-ray number counts from the $Chandra$ Deep Field
North (CDF-N) and South (CDF-S) and found that $\sim 90 \%$ ($\sim 93 \%$) 
of the CXRB is resolved into discrete X-ray sources in the 0.5-2 keV (2-8 keV) band.
Using the $XMM$-$Newton$ observation of the Lockman Hole and the CDFs,
\citet{wor05} found that the resolved fractions of the CXRB are $\sim85\%$ (0.5-2 keV),
$\sim80\%$ (2-10 keV), and $\sim50\%$ ($\gtrsim 8$ keV), respectively.  
Therefore, the CXRB is predominantly resolved into discrete
sources in the 0.5-2 keV
and 2-10 keV bands; however, the constituents of the CXRB in the hard energy band 
($\gtrsim 8$ keV) are still unknown.
 
The deepest X-ray surveys are 
the $2Msec$ CDF-N \citep{bra01,ale03} and
the $1Msec$ CDF-S \citep{gia01,ros02} covering
small sky areas ($\sim0.12~deg^2$).
The faint flux limits of the CDFs are
$\sim2\times10^{-17}~erg~cm^{-2}~sec^{-1}$ (0.5-2 keV) and
$\sim2\times10^{-16}~erg~cm^{-2}~sec^{-1}$ (2-8 keV), respectively.
The $XMM$-$Newton$ survey of the Lockman Hole covers $\sim0.43~deg^2$ 
with an effective exposure time of $\sim700$ $ksec$ 
(flux range of $\sim few~ \times 10^{-16} \sim3-5\times10^{-14}$
$erg~cm^{-2}~sec^{-1}$ in each band) \citep{wor04}.
The XMM large scale structure survey ($XMM$-$LSS$) is a medium depth ($\sim10~ksec$)
and large area ($\sim64~deg^2$) X-ray survey \citep{pie04}.
The $3~deg^{2}$ XMM medium depth survey (XMDS)
pointed at the center of $XMM-LSS$
reaches a flux limit of
$\sim10^{-15}~erg~cm^{-2}~sec^{-1}$ in the 0.5-2 keV band \citep{chi05}.
The $Chandra$ X-ray survey of the NDWFS Bo$\rm{\ddot{o}}$tes is a
contiguous wide ($\sim 9.3~deg^{2}$) and medium depth ($\sim 5~ksec$)
survey and contains 4642 sources with flux limits of
$4\times10^{-15}$ $erg~cm^{-2}~sec^{-1}$ in the 0.5-7 keV band \citep{mur05}.

The Chandra Multiwavelength Project (ChaMP) is a serendipitous
$Chandra$ archival survey of X-ray sources covering a wide area ($\sim 10~deg^{2}$)
with a range of depths $0.9\sim124~ksec$ exposures. 
The main scientific goals of the ChaMP are to investigate the 
(1) formation and evolution of high redshift AGN,
(2) properties of X-ray luminous galaxies and clusters,
and (3) constituents of the CXRB.
Kim, D.-W. et al. (2004a, hereafter Paper I) reported
the first ChaMP X-ray source catalog 
including 991 near on-axis, bright X-ray sources obtained from
62 ChaMP fields and
\citet{gre04} have performed optical and spectroscopic follow up 
observations of a subset of these ChaMP X-ray sources. 

Using the first ChaMP X-ray point source catalog and the follow up
surveys of optical and spectroscopic observations, 
there have been several interesting results.
\citet{kim04b} established the number counts of the
ChaMP X-ray point sources in the 0.5-2 keV and the 2-8 keV bands, which 
agreed with previous studies within the uncertainties, and found that there are no 
significant field-to-field variations in cosmic X-ray source number density
on the scale of $\sim 16 \arcmin$ which corresponds to a single  
$Chandra$ observational field of view.
\citet{sil05a} found the turnover in the co-moving space
density of X-ray selected, luminous type 1 AGN ($log L_{x} > 44.5~erg~sec^{-1}$
measured in the 0.3-8 keV band) to be at $z>2.5$ consistent with the optical results.
The hard X-ray emitting AGNs in twenty ChaMP fields were investigated
and classified as broad emission-line AGN ($62\%$), narrow emission-line
galaxies ($24\%$), absorption line galaxies ($7\%$), stars ($5\%$),
or clusters ($2\%$). 
Most X-ray unabsorbed AGN ($N_{H}<10^{22}~cm^{-2}$) 
have broad emission lines and blue optical colors but there is  
a significant population of redder AGNs with broad optical
emission lines.
Most X-ray absorbed AGN ($10^{22}<N_{H}<10^{24}~cm^{-2}$)
are associated with narrow emission-line galaxies, those with red optical 
colors being dominated by luminous, early type galaxy hosts 
rather than dust reddened AGN \citep{sil05b}.
\citet{bar06} presented the ChaMP X-ray extended source catalog
which contains 55 extended sources from 130 $Chandra$ fields. From the
overlapping optical/X-ray fields ($6.1$ $deg^{2}$) they found 115 optical
cluster candidates of which 13 were detected as extended X-ray sources.
A comparison of the richness of the optical-only versus X-ray/optically
matched cluster samples shows that the average richness of the optical-only
clusters is smaller by $\sim 4\sigma$ than the matched X-ray/optical
clusters. This result suggests that the optical-only sample is either
(1) composed of mainly poor systems that lack sufficient hot gas for
detection in the X-rays, or (2) are contaminated by nonvirialized filaments
associated with the large-scale structure.
\citet{kim06c} investigated the normal galaxies at intermediate redshift
in the ChaMP fields and found that normal galaxies at redshift, 
$z<0.1$ do not show significant evolution in $L_X/L_B$. They built cumulative 
number counts and luminosity functions of the normal galaxies, and they found
that a group of NELGs appear to be heavily obscured in X-rays while 
the low redshift AGNs in this sample do not appear to be significantly absorbed.
Also, they found two E+A galaxy candidates and they concluded that those
galaxies support the merger/interaction scenario of galaxy formation
from their X-ray spectra studies.

In this paper, we present the ChaMP X-ray point source catalog including
$\sim 6,800$ X-ray point sources obtained from 149 ChaMP fields
and covering a sky area of $\sim 10~deg^{2}$. 
Compared to the first ChaMP X-ray point source catalog, this catalog contains
seven times more sources, covers three times more
sky area, and includes fainter sources and those with larger off axis angle.
We performed extensive simulations to investigate the sensitivity, source
probability, and positional uncertainty of the ChaMP sources. 
This catalog allows more statistically robust 
results from X-ray point source studies.  
The ChaMP data reduction procedures are similar to those in
Paper I, therefore, we briefly summarize and/or skip those parts which are already
described in Paper I and concentrate on newly added or improved procedures. 
In \S2, we summarize the selection criteria 
and properties of the ChaMP fields. In \S3, the data reduction and analysis 
of the ChaMP are described.
\S4 discusses the process and results of the ChaMP simulations.
The ChaMP X-ray point source catalogs are provided in \S5, and 
a summary and conclusions are given in \S6.  

\section{ChaMP Field Selection}
We selected $Chandra$ fields observed with ACIS at high
Galactic latitude, $|b|>20^{\circ}$ and excluded those fields
containing large extended sources, planetary observations, and local
group galaxies. Fields intended by their PIs as surveys were also excluded
(see Paper I). These selection criteria yield 149 ChaMP fields in 
$Chandra$ cycles 1 and 2, consisting of 35 ACIS-I and
114 ACIS-S observations. 7 ACIS-I and 28 ACIS-S ChaMP fields 
partly overlap one another on the sky,
and those sources detected in multiple observations
are listed separately, for example to allow study of source variability.  In
Table \ref{tbl-champ-list}, the observational parameters of 149 ChaMP
fields are listed in order of right ascension.

In Figure \ref{fig-pos}, we display the 149 ChaMP field locations
in equatorial coordinates. Red circles represent ACIS-I at the
aim point, and blue circles ACIS-S. 
The circle size crudely indicates the Chandra exposure time,
ranging from 0.9 to 124 ksecs.
The ChaMP samples are uniformly distributed over the
entire celestial sphere except (by selection) the Galactic plane region.
Figure \ref{fig-expnh} shows the number distributions of the exposure times
and Galactic extinction of the ChaMP fields in top
and bottom panels, respectively. The mean exposure time of the
ChaMP is $\sim25~ksec$ and the mean Galactic extinction, 
$N_{H}=(3.4\pm2.2) \times 10^{20}~cm^{-2}$. The ChaMP
samples cover a wide range of exposure times and the Galactic
extinction of the ChaMP fields are generally much lower than those of
Galactic plane ($N_{H}\sim 10^{22}\sim10^{23}~cm^{-2}$). The 62 ChaMP
fields included in Paper I are represented by shaded histograms. In
this study, the X-ray point source catalog includes all X-ray sources
in 149 ChaMP fields as well as fainter and larger off axis angle
sources, while the catalog in Paper I included only near
on-axis ($off$ $axis$ $angle$ $<6\arcmin$ 
or S3 chip for ACIS-S observations) and 
bright ($net~counts>20$) sources in 62 ChaMP fields.

\section{ChaMP Data Reduction}
We have developed a ChaMP-specific pipeline, XPIPE, to reduce the
$Chandra$ data. The pipeline consists of three main parts; (1) data
correction and data screening using the CIAO\footnote {See
http://cxc.harvard.edu/ciao.} package, (2) source detection using
the $wavdetect$ tool in the CIAO package, and (3) source extraction using
the $xapphot$ tool \citep{kim06b,mar06} based on $cfitsio$\footnote {See
http://heasarc.gsfc.nasa.gov/docs/software/fitsio/fitsio.html.}
library. The data correction and data screening procedures are the
same as in Paper I. We do not use sources detected in the S4 chip
(CCDID=8) because of the streaking problem (see Paper I).

\subsection{Source Detection}
For source detection, we use the $wavdetect$ tool available in the
CIAO package. $wavdetect$ consists of two parts: 
$wtransform$, convolving the data with the wavelet function for
selected size scales; and $wrecon$, constructing a final
source list and estimating various parameters for each source
\citep{fre02}. We run $wavdetect$ in the B band (see Table
\ref{tbl-ebands} for energy band definition)
with a significance
threshold parameter of $10^{-6}$, which corresponds to one possible
spurious pixel in one CCD (see \S 4.3.2 for our simulation
results on the probability of finding a spurious source). 
We select a range of scale size parameters in
seven steps from 1 to 64 pixels ($1
pixel=0.492\arcsec$). 
To avoid finding spurious sources located at the edge of
the CCD chips, 
a minimum of $10\%$ of the on-axis exposure was required for source
detection. 
Exposure maps of the ChaMP fields were generated 
for each CCD at an energy of
1.5 keV with an appropriate aspect histogram
\footnote{See
http://cxc.harvard.edu/ciao/threads/expmap\_acis\_single.}.  
Other parameters were set at the default values given in $wavdetect$.
The positions provided by $wavdetect$ in CIAO 2.3 for
off-axis sources, where the shape of the Point Spread Function 
(PSF) is highly asymmetrical and the background contribution to the 
counts in the source cell is non-negligible,
are less accurate than the positions of on-axis sources (Paper I).
To alleviate this problem, we applied a position refinement algorithm
(P. Freeman 2003, private communication),
which iteratively redetermines  
the position of the off-axis source until it converges on the best centroid
(see Paper I for detail descriptions),
to the X-ray positions determined
by $wavdetect$ in CIAO 2.3. 
When CIAO 3.0 or later versions are used,
this process is not necessary, because 
the position refinement algorithm 
has been applied in $wavdetect$\footnote{See
http://cxc.harvard.edu/ciao/releasenotes/ciao\_3.0\_release.html.}.

The size and shape of the PSF for $Chandra$  
varies as a function of off axis angle and radial
direction. $wavdetect$ 
uses a Mexican Hat function, a reasonable function 
for mirrors/detectors which are characterized by a quasi-Gaussian PSF, and
which detects sources successfully in most cases. 
However, the off-axis PSF is asymmetric
and contains sub-structure in the core, causing
$wavdetect$ to detect sometimes a spurious pair of double sources.
We note that the sub-structure of the off-axis PSF can be resolved  
by $wavdetect$ because the size of the PSF becomes larger with 
increasing off axis angle.  
This PSF effect can be corrected by a PSF deconvolution (Paper I).
In Paper I, source pairs with small separations were inspected and 
three pairs of spurious doubles were found, bright
($net~counts >$ a few hundreds counts)
enough to deconvolve with their PSFs.
In this study, to identify and correct this PSF effect, 
we generated a single PSF image at the median
location of each overlapping pair of sources 
whose positional centers are very close together 
(see \S 3.2.2 for quantitative definition of large
overlapping sources)
and a second image using a combination of two source PSFs.
We then compared the observed source image with the modeled images.
To generate the PSF images, we used a PSF ray trace tool $ChaRT$\footnote{See 
http://asc.harvard.edu/chart/threads/index.html.} and 
a $Chandra$ detector simulation tool MARX\footnote{See http://space.mit.edu/CXC/MARX/.}
assuming a monochromatic energy at 1.5 keV and a source counts ratio
corresponding to that of overlapping sources as determined by XPIPE.

We found twelve pairs of spurious double sources and
Figure \ref{fig-psfeffect} shows the observed sources image and 
the modeled images for a sample pair.
First, we can see that the PSF shape at this location is
asymmetric ($middle$) and the positional centers of the spurious double 
sources are located along the elongated PSF direction.
The shape of the observed X-ray sources ($left$) is similar to the single
PSF ($middle$) rather than that of the double source PSFs ($right$).
Therefore, we conclude that $wavdetect$ has incorrectly detected 
a single source as a double source due to the asymmetric sub-structure in the PSF.
All double sources found to be spurious are too faint 
($net~counts <$ a few hundred counts) to deconvolve with their PSFs.
Therefore, we assigned the median position of the spurious double sources as the
new position of the single source and half of the distance 
between double sources
was quadratically summed to their positional uncertainties
(see \S 4.2.1 for the positional uncertainty of the ChaMP X-ray point sources).
The source counts were then extracted at the new source position
(see \S 3.2 for the source counts extraction).

\subsection{Source Properties}
After detecting X-ray sources with $wavdetect$, we extracted their X-ray
properties by applying aperture photometry. Since $wavdetect$
sometimes underestimates the net counts for faint sources (see \S
4.3), we do not use the $wavdetect$-determined X-ray photometry.
Instead, we apply an aperture photometry source extraction tool $xapphot$, 
developed for a general purpose and
applicable to both $Chandra$ and XMM-Newton data \citep{kim06b}.
We note that  
XPIPE detects the source positions only in the B band 
with $wavdetect$ and 
applies the same position and size for the source extraction regions 
in every energy band. 
The reliability of the ChaMP source properties using 
XPIPE will be discussed in \S 4.3.

\subsubsection{Source Count Extraction Regions}
With the X-ray source position determined by $wavdetect$, we extract
source counts from a circle with a $95\%$ encircled
energy radius, determined at 1.5 keV from the PSF table
\footnote{See http://cxc.harvard.edu/cal/Hrma/psf/index.html.}.
A minimum radius of $3\arcsec$ and maximum of $40\arcsec$
are chosen to avoid small number statistics in the source counts and
severe fluctuations in the background sky. The source radii in this
study are slightly smaller than those in Paper I which used an older
version of the PSF tables. In the top panel of Figure \ref{fig-prop}, 
the source radius for Paper I ($dotted$ $line$) 
and that for this study ($solid$ $line$) are
displayed as a function of off axis angle. The difference between
source sizes is plotted as a dashed line, indicating a significant
difference at large off axis angle. 
The source size is reduced by a maximum of
$\sim18\arcsec$ at off axis angle of $\sim13\arcmin$ and 
unchanged in the axis region of $\lesssim 3 \arcmin$ compared to
the old source size. In the bottom panel of Figure \ref{fig-prop}, the
difference between source counts in this and previous studies are
displayed. The reduced source radii yield an average net counts lower by 
$2\pm7\%$ in this catalog compared with Paper I.

The size of the background extraction annulus is a free parameter 
generally chosen within the range 2 to 5 times the source radius,
depending on local and global background fluctuations.   
However, for point sources inside an extended source, the
size of the background extraction annulus was set to 1 to 2 times 
the source radius because in this case
local variations in the
background are much more important than global variations. The ChaMP
X-ray extended sources are identified by $wavdetect$ with a large
wavelet and source properties extracted via fitting with a Gaussian
profile and a $\beta$ model. The ChaMP extended source catalog used in
the ChaMP point source photometry is provided in a separate paper
\citep{bar06}.

\subsubsection{Net Counts}
The net counts $N$ of a source in a given energy band are determined by
subtracting the normalized background counts from the source counts in
the source region as follows:
\begin{equation}
N=N_{S}-N_{B}/AR,
\end{equation}
where $N_{S}$ and $N_B$ are the total counts in the source and
background regions, respectively. 
The normalization factor $AR$ in equation (1) is
given by:
\begin{equation}
AR={<E_{B}> A_{B} \over <E_{S}> A_{S}},
\end{equation}
where $<E_S>$ and $<E_B>$ are the mean exposure times for the source
and background regions, respectively, and $A_S$ and $A_B$ are the
geometric areas of the two regions. To avoid 
contamination in the background region, we exclude other point and
extended source regions within the background region.
The net counts errors are derived following \citet{geh86}.

There is a significant probability that two or more sources will overlap
with each other, especially at large off axis angle.
Note that the PSF size increases exponentially with increasing
off axis angle. For overlapping
sources, simple aperture photometry overestimates the
source counts.  While simultaneous fitting of multiple
PSFs may be a good way to deconvolve
overlapping sources, this process
requires sufficient counts (a few hundred), which is
unusual for typical X-ray observations. Thus to determine the net
counts for overlapping sources, we apply two independent correction
methods as in Paper I, a small overlap correction and a large overlap
correction, depending on the amount of overlapping area involved.

In the left side of Figure \ref{fig-small_large}, we display a
schematic diagram of a small overlapping source, in which the
distance between them $D_{12}$ is greater than the radius of each source
but less than the sum of their radii: 
\begin{equation}
0.5< {{D_{12}} \over {D_{PSF}}} < 1,
\end{equation}
where $D_{PSF}$ is the $95\%$ encircled energy diameter of the PSF.
A small overlap is
the most common overlap among X-ray sources due to their 
relatively low density in the ChaMP fields.
The corrected net counts $N_{1}$ and $N_{2}$ 
for the overlapping sources S1 and S2
are estimated as follows:
\begin{eqnarray} 
N_1 & = & 2\pi N_{1,A_1}/(2\pi-\theta_1), \\
N_2 & = & 2\pi N_{2,A_2}/(2\pi-\theta_2),
\end{eqnarray} 
where $N_{1,A_1}$ and $N_{2,A_2}$ are the net counts of S1 in area
$A_1$ and the net counts of S2 in area $A_2$, respectively.
$\theta_{1}$ and $\theta_{2}$ are the angles in units of radians 
covered by sectors $B_{1}$ and $B_{2}$ of the overlapping sources, respectively. 
We assumed a radially symmetric event distribution for both X-ray sources. 

The right side of Figure \ref{fig-small_large} displays an example of large
overlapping sources, where the distance $D_{12}$ is less than the radius of each source.
In this case, the center of each source is located within the source region of the 
overlapping partner, 
such that the above algorithm is not applicable.
To correct this large overlapping case, first, we defined
the core radius $R_{c}$ of each source as follows:
\begin{equation}
R_{c} \equiv max\left({D_{12} \over 3},~2~pixels\right),
\end{equation}
where $D_{12}$ is the distance between two overlapping sources. 
$R_{c}$ has a minimum of $2~pixels$ to allow a statistically robust estimation
of the counts within the core radius. The corrected net counts of the large overlapping 
sources $N_{1}$ and $N_{2}$ are estimated as follows: 
\begin{eqnarray}
N_1 = N_t (2N_{1,A_1} + N_{1,C_1})/N_0, \\
N_2 = N_t (2N_{2,A_2} + N_{2,C_2})/N_0, \\
N_0 = (2N_{1,A_1} + N_{1,C_1}) + (2N_{2,A_2} + N_{2,C_2}),
\end{eqnarray}
where $N_t$ is the sum of net counts of S1 and S2
(i.e. net counts in a union area of two source regions).
$N_{1,A_1}$ and $N_{2,A_2}$ are the net counts of S1 in region
$A_1$ and the net counts of S2 in region $A_2$, respectively.
$N_{1,C_1}$ and $N_{2,C_2}$ are the net counts of S1 in core region
$C_1$ and the net counts of S2 in core region $C_2$, respectively.
The radii of core regions $C_1$ and $C_2$ are calculated with equation (6) and have
the same size.
Since the core region of each source is contaminated by photons from the overlapping
source, the photons in unperturbed regions 
($A_1$ and $A_2$) are weighted higher than those in core regions ($C_1$ and $C_2$)
by a factor of two.
Excluding spurious sources, small and large overlap corrections were applied
to $2.5\%$ and $0.6\%$ of the ChaMP sources, respectively.
We note that $xapphot$ does not include the correction procedure
for a source overlapping largely with more than one source.
However, only one such case is included in the ChaMP X-ray point source catalog:
CXOMP J111816.9+074558, CXOMP J111816.8+074600, and CXOMP J111816.8+074557  
overlap largely each other, and they are the target of the observation 
(OBSID=363, gravitationally lensed quasar) having a pile-up flag 
(flag=37, see Table \ref{tbl-flag} in \S 3.2.5).

\subsubsection{Hardness Ratio and Colors}
The X-ray point source properties are extracted in 
the five ChaMP specified energy bands and in the three 
commonly used energy bands.
The used energy bands and definitions of hardness ratio (HR) 
and X-ray colors (C21 and C32)
are listed in Table \ref{tbl-ebands}
and their scientific rationale was described in Paper I.
The HR and X-ray colors can be calculated from the source net counts
in two different energy bands according to their definitions 
(hereafter classical method).
However, for the faint sources,
the HR, C21, and C32 and their error propagations from the classical
method are often unreliable or unrealistic
because of negative/undetectable net counts in one band or a non-Gaussian
nature.
Therefore, we calculated the HR, C21, and C32
with a Bayesian approach which models the detected
counts as a Poisson distribution 
and which gives reliable HR and X-ray colors  
for both low and high count sources \citep{van04,par06}. 

To derive the HR and X-ray colors with the Bayesian approach, we used the 
BEHR\footnote{See http://hea-www.harvard.edu/AstroStat/BEHR/.} 
program (version 07-27-2006, Park et al. 2006) with the required inputs: 
source counts, background counts, and
ratio of background area to source area in both energy bands.
We assumed a non-informative, flat, prior distribution on the linear scale
($softidx=hardidx=1$). 
We note that the energy-dependent vignetting in the soft and hard counts
is not corrected for deriving the HR.
The BEHR program calculates the solution with two different method:
a Gibbs sampler (Monte Carlo integration) and a Gaussian quadrature
(numerical integration). The Gibbs sampler is efficient but less accurate
than the Gaussian quadrature for faint sources; however, the 
Gaussian quadrature becomes less efficient with increasing source counts.
Therefore, we used the Gibbs sampler for bright sources ($net~counts>15$
in two energy bands) and Gaussian quadrature for faint sources
($net~counts<15$ in at least one energy band), respectively.
The default values were used for the remaining optional inputs.  
The BEHR program calculates the mode, mean, and median of the 
posterior probability
distribution. The mean of the distribution is 
a robust estimator for the HR, while the mode for the X-ray colors \citep{par06}. 
In Figure \ref{fig-hcolor_final}, we compare the classical method with
the Bayesian approach for HR and X-ray colors.
For bright sources ($S/N>2$), the HR and X-ray colors from both methods 
agree well; however, for faint sources ($S/N<2$), they 
do not agree, because the classical method using the Gaussian statistics 
fails to describe the nature of faint sources.

\subsubsection{Source Flux}
\begin{figure}
\epsfig{figure=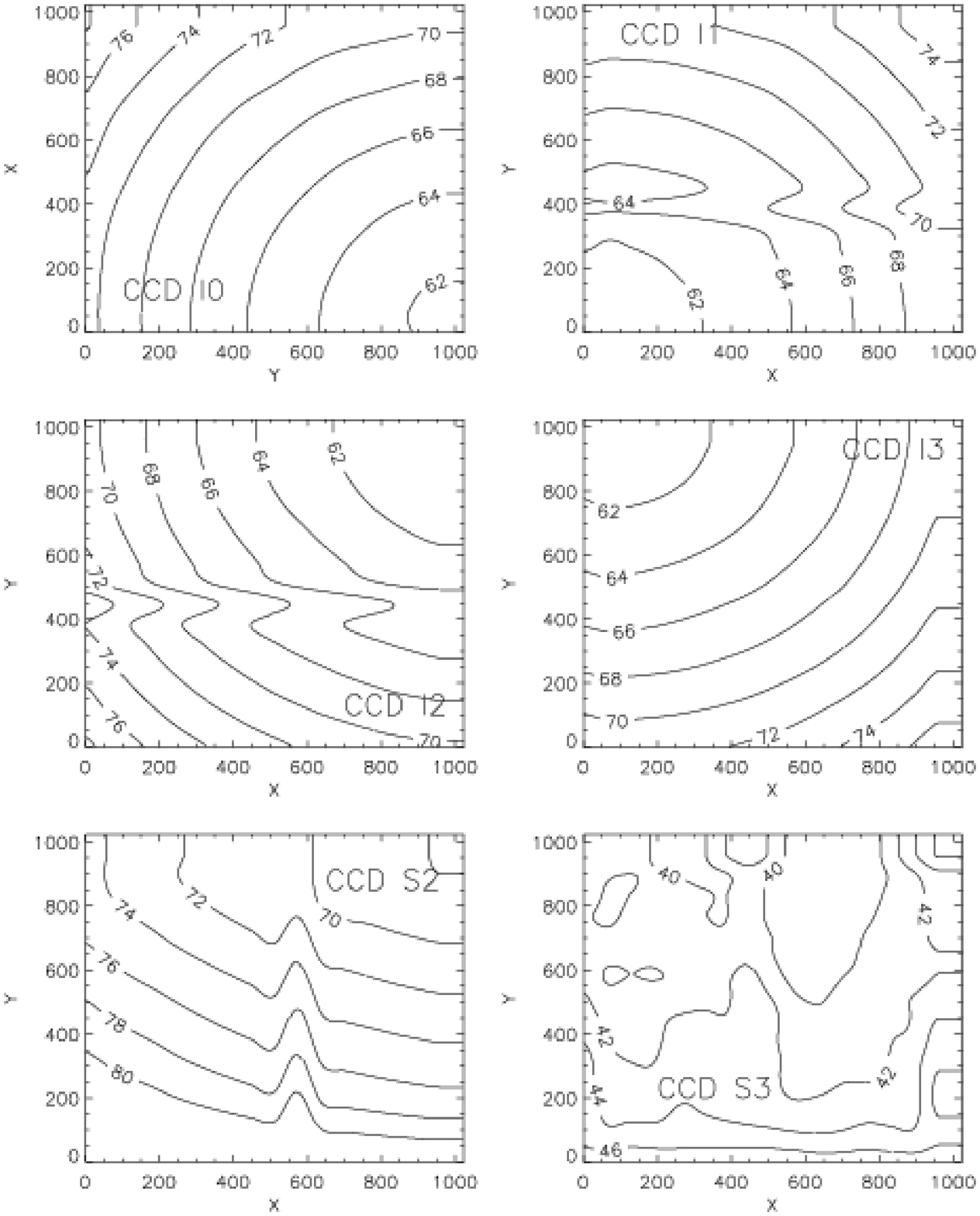, height=0.975\textwidth,width=0.92\textwidth}
\caption{
Contour maps of the energy conversion factors ($ECF$) in ACIS-I 
observations including ACIS-S S2 and S3 CCD chips. 
Figures are shown in chip coordinates rotated by $90^{\circ}$ 
(CCD I1 and I3) and $270^{\circ}$ (CCD I0 and I2) relative to CCD S3.
The energy band is 0.3-2.5 keV and
a photon index of $\Gamma_{ph}=1.7$ was assumed. 
The $ECF$s are calculated at $16\times16$ grid points with a grid size of
32 pixels and smoothed with a cubic kernel. 
The $ECF$ is in units of $erg$ $cm^{-2}$ $count^{-1}$.
}
\label{fig-ecf_vari_con}
\end{figure}

In general, the $Chandra$ X-ray source flux is determined as follows:
\begin{equation}
Flux=count~rate \times ECF, 
\end{equation}
where the $ECF$ 
is the energy conversion factor which converts source count rate to 
source flux in units of 
$erg$ $cm^{-2}$ $count^{-1}$.
The $ECF$ varies with observation date and CCD pixel position
because of the temporal and spatial variations of the ACIS CCD quantum efficiency 
\footnote{See
http://asc.harvard.edu/cal/Acis/Cal\_prods/qeDeg 
for the low energy QE degradation.} and the vignetting effect. 
The temporal QE variation of the $ECF$ can be corrected by generating
$ECF$s per observation and per CCD chip. 
To investigate the spatial variation of the $ECF$,
we generated the 0.3-2.5 keV $ECF$ map of an ACIS-I
observation including ACIS-S S2 and S3 CCD chips. 
Using the redistribution matrix function (RMF)
and ancillary response function (ARF) files 
and assuming a photon index of $\Gamma_{ph}=1.7$ and Galactic absorption 
$N_H$ \citep{sta92} for a given observation,
we derive $ECF$s with $Sherpa$
\footnote{See http://asc.harvard.edu/sherpa/threads/index.html.}
in $16\times16$ grid points with grid size of
32 pixels in each CCD chip (here after $ECF_{grid}$).
In Figure \ref{fig-ecf_vari_con}, we display the
$ECF_{grid}$ contour maps smoothed with a cubic kernel 
and the left panels of Figure \ref{fig-ecf_cor_vari} shows
the $ECF_{grid}$ as a function of off axis angle in each CCD chip. 
The $ECF$ spatially varies by up to $\sim 25\%$.

To quantitatively see the spatial variation of the QE, 
we display the ratios of the $ECF_{grid}$ over
the $V_{cor} \times$ $ECF_{single}$ as a function of the off axis angle 
in the right panels of Figure \ref{fig-ecf_cor_vari}, where
the $ECF_{single}$ is the $ECF$ calculated at a single position 
(the aim point position for ACIS-I CCDs and at the maximum exposure positions
for S2 and S3 chips) and
$V_{cor}$ is the vignetting correction factor which is estimated 
from the exposure map at each grid position.
The vignetting corrected $ECF_{single}$ agrees well with $ECF_{grid}$
with the exception of points that are estimated from the CCD 
edge and bad pixels/columns 
($blue$ $squares$). 
The spatial variation of the QE is shown at large off axis angle 
($red$ $triangles$);
however, the deviation is less than $5\%$.
Therefore, in this study, we ignore the spatial variation of the QE and
correct the vignetting effect to determine the source flux as follows:
\begin{equation}
Flux=count~rate \times ECF_{single} \times V_{cor}.
\end{equation}
In Table \ref{tbl-energy-con-factor}, the $ECF_{single}$ are listed 
per observation (OBSID) and per CCD chip and 
calculated at the aim point for I0-I3 of ACIS-I observation
and S3 of ACIS-S observation. For the remaining chips, $ECF_{single}$
is calculated at the maximum exposure position.
For general usage, we calculated $ECF_{single}$ assuming 
various photon indices, $\Gamma_{ph}=1.2$, 1.4, and 1.7
and Galactic absorption $N_H$ from \citet{sta92} for that observation.
We provide the effective exposure time of each X-ray source 
corrected for the vignetting effect at the source position
(see Table \ref{tbl-main-list} and
\ref{tbl-sub-list} in \S5.1). 

\subsubsection{Source Flags}
All X-ray sources in the ChaMP catalog have been visually inspected 
to flag those sources with various special
issues, as listed in Table \ref{tbl-flag}.
Flags 11 to 51, 53, and 54  were determined only by visual
examination.
The spurious double sources due to PSF effects (flag=15 and 38) are
described in detail in \S 3.1.
Since 35 of the 149 ChaMP fields partly overlap on the sky
as seen in the $11^{th}$ column of Table \ref{tbl-champ-list}, 
453 sources were likely
observed more than once (flag=52) in these overlapping fields.
We identified these 453 source candidates by their positions,
matching sources in multiply observed fields within a
$95\%$ confidence level positional uncertainty
(see equation (12) in \S 4.2.1).
False sources having flags from 11 to 21 and  
extended X-ray sources (flag=51) including the X-ray jets (flag=54)
are not listed in the ChaMP X-ray point source catalogs.

To remove bad pixels/columns, we used the bad pixel file. Additional  
hot pixels and bad columns were identified by visually inspecting each CCD  
image and an event histogram as a function of chip x-coordinate (see 
\S 3.1 in Paper I for details). The bad pixels are then included in  
generating an exposure map, which is in turn used to calculate the count  
rate and flux. Although the source flux may be slightly underestimated  
when a bad pixel sits at a source location, the effect of a single  
bad pixel is considerably mitigated by the aspect dither (following a  
Lissajous pattern over 16x16 $arcsec^{2}$).
Therefore, we flagged sources within which a bad pixel/column exists 
as flag=31 following visual inspection.

When the source is located at the edge of the CCD chip, 
where the minimum exposure value
in the source region is less than $10\%$ of the maximum exposure value,
flag 61 is assigned.
The edge flag 61 and overlapping flags from 62 to 68 are 
automatically flagged by $xapphot$.
The overlapping flags correspond to flags $32-35$ in Paper I,
with more detailed classes included here: the overlapping class is
subclassified as either small or large overlaps (see \S 3.2.2 for the
definition of small and large overlap).

\section{ChaMP X-ray Point Source Simulations}
\subsection{Simulation Procedure} 
To investigate source reliability and sensitivity, 
and to establish the empirical equations for 
positional uncertainty on X-ray sources in the ChaMP fields, we have performed
extensive simulations.
The technique was based on that of \citet{kim03} and
consists of three parts, (1) generating artificial X-ray sources with
MARX\footnote{See http://space.mit.edu/CXC/MARX/ and MARX 4.0 Technical Manual.},
(2) adding them to the observed image, and (3) detecting
these artificial sources with $wavdetect$ and extracting source
properties with the $xapphot$. 
We have used every observed ChaMP field for our simulations, 
rather than blank background sky fields, 
to investigate the effects of background counts 
and source confusion. 

We used the active I0, I1, I2, and I3 CCD chips for ACIS-I, and I2, I3, S2,
and S3 CCD chips for ACIS-S $Chandra$ observations, and simulated 
1,000 artificial X-ray sources per $Chandra$ observation. 
The number of detected artificial sources in each field depends on 
the effective exposure time and 
the observed region 
of the sky with various values of $N_{H}$.
On average, $11.4\%$ of the 146,178 input artificial 
X-ray sources are detected in our simulations,
for a total $16,676$ artificial X-ray sources in 149 ChaMP fields.
The number of detected artificial X-ray sources is 2.3 times the
$7,106$ ChaMP sources in the same CCD chips and observations 
and statistically sufficient 
to estimate the properties of the ChaMP fields.
 
The form of the assumed number counts distribution
is not critical to determine the detection probability,
which is determined by the ratio of input to output numbers
at a given flux \citep{vik95,kim03}.
The actual X-ray differential number counts are described by a
broken/double power law
with faint and bright slopes of $\sim-1.5$ and $\sim-2.5$, respectively,
\citep{yan04,bas05,chi05}
in most energy bands; however, the break flux has not been well determined.
Therefore, we assumed a cumulative number counts distribution with a
single power law and a slope of $-1$ corresponding to
a slope of $-2$ in the differential number counts,
taking the average of the faint and bright slopes
from the literature, in the 0.3-8 keV band.
The flux of an artificial source was randomly selected from the
assumed number counts distribution in a flux range of  
$5\times10^{-16}$
$-$ $5\times10^{-10}$ $erg~cm^{-2}~sec^{-1}$,
and the MARX generated the artificial sources with a flux range of
$8\times10^{-17}$ $-$ $2\times10^{-11}$ $erg~cm^{-2}~sec^{-1}$ 
including the Poisson uncertainty of
the input source counts.
The flux range of the detected artificial sources spans 
$1\times10^{-17}$ $-$ $2\times10^{-11}$ $erg~cm^{-2}~sec^{-1}$
which covers the flux
range of the actual ChaMP X-ray point sources, $1\times10^{-17}$
$-$ $6\times10^{-12}$ $erg~cm^{-2}~sec^{-1}$. 

We assume a power law spectrum
with a photon index of $\Gamma_{ph}=1.7$, because the  
ChaMP X-ray points sources with $S/N>1.5$ have $\Gamma_{ph}=1.5\sim2$ 
(Kim, D.-W. et al. 2004b; 
see Figure \ref{fig-color-acisi} and \ref{fig-color-aciss}
in this paper). 
We note that these sources cover a flux range of
$4\times10^{-16}\sim2\times10^{-12}$ (0.5-2 keV) and
$2\times10^{-15}\sim7\times10^{-12}$ (2-8 keV) in $erg$ $cm^{-2}$ $sec^{-1}$, 
respectively.
\citet{toz06} performed X-ray spectral analysis for 82 X-ray bright
sources in the CDF-S, and they found that the weighted mean value for
the slope of the power law spectrum is $<\Gamma_{ph}>\simeq1.75\pm0.02$.
The flux range of these bright sources in the CDF-S overlaps with
the faint flux end of the ChaMP sources, therefore, 
we assumed that the faint ChaMP sources ($S/N<1.5$) also have a photon index 
of $\Gamma_{ph}\sim1.7$.
We assumed Galactic absorption,
$N_{H}$, \citep{sta92} for each observation; however, did not include 
intrinsic absorption in the artificial source spectrum.
The spectrum of each X-ray point source was generated 
using the XSPEC\footnote{See http://xspec.gsfc.nasa.gov/.} package. 

The position of an artificial source was randomly selected on each CCD chip,
but it was rejected if the source area at a given random position had an 
exposure map value with less than $10\%$ of the maximum.
This requirement is identical to that 
in the ChaMP X-ray point 
source reduction procedure.
To avoid over-crowding of the artificial sources, 
$\sim250$ artificial sources per CCD were 
divided into several groups to be added into the observed image:  
while we did not allow the artificial X-ray point sources to overlap
one another,
we allowed overlap between artificial and real X-ray
sources to provide an estimate of source confusion in each observed field.
This resulted in $\sim10$ ($\sim20$) simulated images per ACIS-I (ACIS-S) 
CCD, corresponding to $\sim10,500$ CCD images (event files) 
to run through $wavdetect$ ($xapphot$).
Since $\sim11.4\%$ of the artificial sources are detected on average 
we added only $\sim1.5$ artificial sources to each simulated image.
The net counts of the overlapping artificial sources with real sources were
corrected following the overlapping source correction methods described in
\S3.2.2.

To correct the temporal QE degradation of $Chandra$ 
\footnote{See CXC Memo on 2002 July 29
(http://cxc.harvard.edu/cal/Acis/Cal\_prods/qeDeg/index.html).}, we
used the $ECF_{single}$ for each observation, as described in \S 3.2.4.
However, because of the mis-match between calibration data used in
MARX Version 4.0.8 and our analysis, there is a slight difference
in the count-flux conversion. 
Thus we performed a set of test simulations for each CCD chip and observation
to correct this mis-match
and then renormalized the MARX output by as much as 
$10\%$ per CCD chip in each observation. 
After generating and adding artificial X-ray point sources 
into observed X-ray images, 
we detected them and 
extracted their source properties with exactly the same techniques
as used in the ChaMP X-ray point source catalog.

\subsection{Positional Uncertainty}
\subsubsection{Empirical Equation of Positional Uncertainty}
The positional uncertainty of $Chandra$ X-ray sources is a function of
source counts, off axis angle, and background counts. To
estimate the positional uncertainty in the ChaMP
fields, we investigated the offset between input 
and detected position for the artificial sources.
To estimate the positional offsets of
artificial sources, first we excluded the observed X-ray sources in
simulated images to avoid the confusion caused by a mixture of 
observed and artificial X-ray sources.  Second, we
matched the input and detected artificial 
sources within twice the input source radius.   
The nearest object in this matching radius was assigned as a matched pair.
An object with more than one match was assigned as a
pair with the nearest neighbor. We then carefully performed
a visual inspection to reject incorrectly matched sources.
In the top panel of Figure \ref{fig-poserr_off_eq}, 
we display the positional offset of the artificial X-ray sources
split into three source count categories
as a function of off axis angle. 
Since the source position is determined by $wavdetect$,
we used the source counts measured by $wavdetect$
rather than by our aperture photometry $xapphot$.
It appears that the positional offsets
exponentially increase with off axis angle 
and decrease as the source count increases with a power law form.

Applying the exponential function and the power law,
we derive empirical equations for the positional uncertainty of ChaMP
X-ray point sources.
A $95\%$ confidence level, these are: 
\begin{equation}
log PU = \left\{\begin{array}{rr}
0.1145 \times OAA - 0.4958 \times logC + 0.1932, & 0.0000 < logC \leq 2.1393 \\
 & \\
0.0968 \times OAA - 0.2064 \times logC - 0.4260, & 2.1393 < logC \leq 3.3000 \\
\end{array}
\right.
\end{equation}
A $90\%$ confidence level:
\begin{equation}
log PU = \left\{\begin{array}{rr}
0.1142 \times OAA - 0.4839 \times logC + 0.0499, & 0.0000 < logC \leq 2.1336 \\
 & \\
0.0989 \times OAA - 0.2027 \times logC - 0.5500, & 2.1336 < logC \leq 3.3000 \\
\end{array}
\right.
\end{equation}
A $68\%$ confidence level:
\begin{equation}
log PU = \left\{\begin{array}{rr}
0.1137 \times OAA - 0.4600 \times logC - 0.2398, & 0.0000 < logC \leq 2.1227 \\
 & \\
0.1031 \times OAA - 0.1945 \times logC - 0.8034, & 2.1227 < logC \leq 3.3000. \\
\end{array}
\right.
\end{equation}
Here positional uncertainty, $PU$, is in arcseconds, 
and off axis angle, $OAA$, is in arcminutes.
Source counts, $C$, are as extracted by $wavdetct$. 
The above equations are valid for the ChaMP X-ray point sources 
with an off axis angle,  
$OAA\lesssim15\arcmin$ and source counts, $logC\lesssim3.3$.
For ChaMP X-ray point sources located at off axis angle
larger than $15 \arcmin$, 
the positional uncertainties were assigned to be $60 \arcsec$. 
In the bottom panel of Figure \ref{fig-poserr_off_eq}, 
using equation (12)-(14), we display the positional uncertainties
as a function of off axis angle for three different source counts. 
Figure \ref{fig-poserr} shows the number distributions of positional uncertainty of
the ChaMP X-ray point sources from equation (12)-(14).
For $68\%$, $90\%$, and $95\%$ 
confidence level positional uncertainty distributions, 
the medians 
are $0.7\pm0.45\arcsec$, $1.3\pm0.8\arcsec$, 
and $1.8\pm1.1\arcsec$, respectively.

We also investigated the dependence of the positional uncertainties 
on the background counts; however,
it is negligible in the ChaMP sample.
Since we excluded high background regions such as the Galactic plane,
the background counts per unit pixel of ChaMP X-ray sources
are only $3 \times 10^{-4} \sim 9 \times 10^{-2}$ $counts~pixel^{-1}$ 
and there are no significant background fluctuations in these fields.
However, the background effect should be carefully considered to estimate
the positional uncertainties of X-ray sources
when the background fluctuations are severe, such as in the Galactic plane.

\subsubsection{Astrometry}
To ensure accurate absolute positions for the ChaMP X-ray point sources,
we apply the standard $Chandra$ aspect offsets
\footnote{See http://cxc.harvard.edu/cal/ASPECT/.}. For the ChaMP data
set, the magnitude of the mean aspect offset correction 
is $\sim 0.5 \arcsec$, and
the maximum is $\sim 2.7 \arcsec$ (for OBSID=521).
To further check the absolute positional accuracy of our ChaMP X-ray
point sources, we matched the ChaMP X-ray sources with the
SDSS-DR3
\footnote{See http://www.sdss.org/dr3/.} (hereafter SDSS)
optical objects,
for which the absolute positional uncertainty is less than $0.5 \arcsec$.
60 of the 149 ChaMP fields overlap with
the SDSS sky regions. 
Using the $95\%$ confidence level positional uncertainty equation
(equation (12) in \S 4.2.1), over $2\arcsec<radius<12.33\arcsec$,
we searched the SDSS optical candidates of the ChaMP X-ray sources.
The minimum searching radius
is large enough not to miss probable partners, and the maximum searching radius
corresponds to half the mean separation of the SDSS objects,
thereby, reducing the number of randomly matched objects.

Some X-ray sources have more than one SDSS candidate counterpart.
To decide the most appropriate SDSS counterpart for these X-ray sources,
we consider their X-ray and optical properties in addition to
their offsets.  
The X-ray sources have a typical relation between X-ray flux and optical 
magnitude \citep{man03,gre04}.
We applied the normalized distance $D_n$ between ChaMP and SDSS sources with
positional uncertainty $PU$ and
X-ray to optical flux ratio $f_{x}/f_{r}$ as follows:
\begin{eqnarray}
D_n=D/PU, \\
log(f_{x}/f_{r})=Log(f_{x})+5.41+0.4\times m_{r},
\end{eqnarray}
where $D$ is the distance between the ChaMP and
the SDSS counterpart,
$f_{x}$ is the X-ray flux in the Sc band, and $m_{r}$ is the
visual magnitude in the $r$ band. 
First, for objects with one counterpart and with $OAA<6\arcmin$,
we calculated the average of $D_n$ and $log(f_{x}/f_{r})$.
Second, for objects with multiple counterparts, 
we calculated the standard deviations of $D_n$ and $log(f_{x}/f_{r})$  
relative to the average $D_n$ and $log(f_{x}/f_{r})$, and
introduced a likelihood as follows:
\begin{equation}
L=\sqrt{\Delta_{log(f_{x}/f_{r})}^{2}+\Delta_{D_n}^{2}}, 
\end{equation}
where $\Delta_{x} \equiv (x-\bar{x})/\sigma_{x}$.
Finally, we chose the counterpart with the lowest $L$ among the multiple 
counterparts as the most appropriate counterpart.

We also calculated the average $log(f_{x}/f_{r})$ as a function
of optical color and optical size. We classified the SDSS
sources into four groups and recalculated the average of $log(f_{x}/f_{r})$
for each group.
The four groups are resolved (galaxies), unresolved with $u-g<0.6$
(UVX QSOs), unresolved with $g-r>1.2$ (M stars), and all other
unresolved. We then redetermined the SDSS counterpart for 
each X-ray source, after
repeating the above procedures.
Except for 20 of $\sim1,600$ pairs, the matching results were same.
To confirm the matching results, we performed a visual inspection
for all matched objects. 
In the top panel of Figure \ref{fig-aspcor},
the positional offsets between the ChaMP and the SDSS source
are plotted as a function of off axis angle.
The positional offset increases exponentially with off axis angle.
The bottom panel shows the number distribution of the positional offsets between
the matched ChaMP and SDSS sources.
The median positional offset of confirmed $\sim900$ matched sources 
is $0.7\pm0.4\arcsec$.

\subsection{ChaMP Source Reliability}
To understand the source reliability of
the ChaMP point source catalog,
we have investigated the detection probability, 
count recovery rate, 
false source rate, and flux limit of each ChaMP field
using the simulation results. 

\subsubsection{Detection Probability} 
The detection probability is determined by the number ratio of
detected artificial sources to input artificial sources. 
Since the sensitivity of the $Chandra$ CCD chip varies spatially,  
the detection probability is a function of off axis angle as well as 
a function of source counts. 
Figure \ref{fig-detectp_c_off} shows
the detection probability of a source as a function of
the B band counts depending on off axis angle.
The detection probability decreases as the source counts decrease, and
as the off axis angle increases because the
sensitivity of the $Chandra$ CCD chip decreases as the off axis 
angle increases.
Sources with counts $>30$ and off axis angle $<5\arcmin$ are detected
with greater than $95\%$ probability.
We note that the detection probability as a function of flux may vary with  
source spectral shape if the source is quite different from our assumed  
spectral shape ($\Gamma_{ph}=1.7$).

\subsubsection{False Source Rate}
We also investigated the probability of false sources in
the ChaMP catalog. 
With simulated images, we found that false sources are 
$\sim 1\%$ of the total detected sources.
$80\%$ of these spurious sources have counts less than $\sim 30$.
Figure \ref{fig-false_count} shows the false source
detection rate as a function of the B band source counts ($top$)
and off axis angle ($bottom$).
The dashed lines indicate the best linear least square fitting results:
\begin{eqnarray}
False~Source~Rate(C)=-0.05(\pm0.00)C+1.92(\pm0.04), \\
False~Source~Rate(OAA)=0.15(\pm0.02)OAA-0.01(\pm0.18),
\end{eqnarray}
where $OAA$ is in units of arcminutes and $C$ is the source counts extracted
by $wavdetect$.
The false source detection rate increases with decreasing source counts
and increasing off axis angle. To derive equation (18), we used false sources
with $C<40$ and for equation (19), we used all false sources. 

\subsubsection{Counts Recovery Rate}
To confirm the quality of our photometry, we investigated 
the count recovery rate defined by the ratio of the difference between
input counts and  output counts ($C_{out}-C_{in}$) 
to input counts ($C_{in}$).
First, we compared the photometry results from $xapphot$ (aperture photometry) 
and $wavdetect$.
The top panel of Figure \ref{fig-xcompmag1} shows
the count recovery
rate using $xapphot$ ($red$ $circles$) and using $wavdetect$ 
($blue$ $squares$)
as a function of input counts.
The count recovery
rate of $xapphot$ is $96\pm1\%$, regardless of input counts. This is very
close to that expected, given that our choice of the source extraction
radius corresponds to the 95\% encircled energy.
We note that our count recovery rate agrees with that of \citet{toz01} 
in which they applied aperture photometry to the CDF-S sources 
with the source extraction region defined as a circle of radius $R_{s}=2.4\times
FWHM$, where $FWHM$ is modeled from the PSF. 
While the $wavdetect$-determined count recovery rate 
is $96\pm2\%$ for bright sources ($counts$ $>50$), it
underestimates the counts for sources with
counts $< 70$. For example, $wavdetect$ recovers $87\pm2\%$ for
sources with input counts $\lesssim 50$. This is primarily because
$wavdetect$ uses a smaller source extraction radius for
fainter sources. 

Second, we have investigated the count recovery
rate of $xapphot$ depending on the off axis angle in the bottom
panel of Figure \ref{fig-xcompmag1}. The $xapphot$ recovers
source counts well regardless of the off axis angle; 
however, the statistical errors of the count recovery rate increase as
the off axis angle increases, since the background fluctuations affect
off-axis sources more severely than on-axis sources due to the larger
source and background extraction regions.

\subsubsection{Flux Limit and Exposure Time}
Using four CCD chips (I0, I1, I2, and I3 for
ACIS-I observations, and I2, I3, S2, and S3 for ACIS-S observations)
per ChaMP field,
we derived the relation between flux limit and exposure time
of the observation in each energy band.
The detected artificial sources with
$S/N>2.0$ have been 
selected in each CCD chip, and their minimum flux is defined as
the flux limit of that CCD chip.
Figure \ref{fig-fluxlim_cband} shows the flux limits of 
detected artificial sources
in the ChaMP fields 
as a function of the exposure time. 
The best linear least
square fit results of the relation between flux limits and exposure times 
in each energy band are as follows:
\begin{eqnarray}
log F_{limit,B}=-1.04(\pm0.02)\times log ET-12.87(\pm0.03),  \\
log F_{limit,S}=-1.06(\pm0.02)\times log ET-13.10(\pm0.02),  \\
log F_{limit,H}=-1.23(\pm0.03)\times log ET-12.08(\pm0.04),
\end{eqnarray}
\begin{eqnarray}
log F_{limit,Bc}=-1.06(\pm0.02)\times log ET-12.81(\pm0.02),  \\
log F_{limit,Sc}=-1.08(\pm0.02)\times log ET-13.17(\pm0.02),  \\
log F_{limit,Hc}=-1.22(\pm0.03)\times log ET-12.15(\pm0.03),
\end{eqnarray}
where $ET$ is the net exposure time of each CCD chip in units of $ksec$
after excluding any background flares (see Paper I) 
and the flux is estimated assuming a photon index of $\Gamma_{ph}=1.7$.
The scatter of the relation is caused by the varying sensitivity
and detection probability of each CCD chip and OBSID. 
These equations give us a representative 
for the flux limit of X-ray sources depending on their exposure time
in the B, S, H, Bc, Sc, and Hc bands
in the ChaMP fields.

\section{ChaMP X-ray Point Source Catalogs}
\subsection{Catalogs}
We found 7365 X-ray point sources in 149 ChaMP fields, after excluding  
false sources (flag=11-21) 
and sources located close to the CCD chip edges (flag=61) 
(see Table \ref{tbl-flag} for flag definitions).
The 102 target point sources (flag=53) 
are included in the catalog for 
completeness. 
We note that, for scientific analysis, target sources need to be carefully 
handled depending on their own scientific goals because they are not random sources. 
For example, we excluded the target sources 
to determine the X-ray point source number counts \citep{kim04b,kim06a}.

Since 35 of the 149 ChaMP fields partly overlap on the sky
as seen in the $11^{th}$ column of Table \ref{tbl-champ-list}, 
there are sources observed more than once  
in these overlapping fields.
For simplicity and flexible usage, we present the sources 
in the overlapping fields in separate tables: e.g., for the X-ray
number counts research, we used only the main ChaMP catalog to 
derive the sky coverage avoiding complex overlapping fields \citep{kim06a}.  
In the main ChaMP tables (Table \ref{tbl-main-list}, Table \ref{tbl-main-phot},
and Table \ref{tbl-main-color}),
we present all sources in fields observed once and 
those in the overlapping fields 
with the longest exposure time.
In the supplementary ChaMP tables 
(Table \ref{tbl-sub-list}, Table \ref{tbl-sub-phot},
and Table \ref{tbl-sub-color}),
we present the sources in the overlapping fields with shorter exposure times. 
The main ChaMP tables list $6,512$ X-ray point sources in
130 ChaMP fields and the supplementary ChaMP tables 
list 853 sources in 19 ChaMP fields. 
Tables \ref{tbl-main-list} and \ref{tbl-sub-list} contain
the source position, positional uncertainty,
off axis angle, source radius, effective exposure time,
and flag. 
Tables \ref{tbl-main-phot} and \ref{tbl-sub-phot} give
the photometry
of the X-ray point sources in eight X-ray energy bands. 
Tables \ref{tbl-main-color} and \ref{tbl-sub-color} list
the hardness ratio and colors such as C21 and C32 
of the X-ray point sources. 
In Table \ref{tbl-multi-list}, we list the same source candidates 
in the overlapping fields (453 pairs/triples of 926 sources)
with their observation date, source counts, count rates, and positional uncertainties.
We note that these candidates are identified only by their positions,
matching sources in the overlapping fields within a  
$95\%$ confidence level positional uncertainty
(see equation (12) in \S 4.2.1). 
These sources also have a source flag of 52 (see \S 3.2.5).  
Table \ref{tbl-multi-list} allows us to investigate the variability 
of X-ray source brightness. 

The ChaMP source name is given by its right ascension and declination.
We note that the position of some sources in Paper I and this paper could
be slightly different because the position refinement process was applied to
sources with an off axis angle of $>400\arcsec$ in Paper I whereas in this paper 
it was applied to all sources. 
For the sources which were published in previous papers 
\citep{kim04a, gre04, sil05a, sil05b},
we use the published name,
even if the source position has been changed.
The full versions of the tables are available 
in the electronic version of this paper
and also on the ChaMP web site\footnote
{See http://hea-www.cfa.harvard.edu/CHAMP.}, while
we only present a sample of each table in this paper.

\subsection{The ChaMP X-ray Point Sources}
To understand the properties of the ChaMP X-ray point sources, 
we investigated their statistical characteristics.
After eliminating source duplications in the main and supplementary 
ChaMP catalogs,
the ChaMP X-ray point source catalog contains $6,889$ unique
sources.   
To eliminate the source duplication,  
we selected a source having the smallest positional uncertainty among the same
source candidates in Table \ref{tbl-multi-list}: since the positional
uncertainty is a function of off axis angle and source counts (see \S4.2.1), 
this criterion automatically selects a higher quality source.  
With these individual sources, we display the distributions 
of source count, flux, 
off axis angle, and hardness ratio and colors of these individual sources with
their median values in Figure \ref{fig-count} through Figure \ref{fig-hcolor},
respectively.
Sources with signal to noise ratio, $S/N>2.0$ are
displayed with shaded histograms.
Figures \ref{fig-color-acisi} and \ref{fig-color-aciss} 
show the X-ray color-color diagrams of the ChaMP X-ray point sources
in four CCD chips observed with ACIS-I and ACIS-S, respectively. 
The grid in the X-ray color-color diagram indicates the predicted
locations of sources at redshift $z=0$ with various photon indices
($0 \leq \Gamma_{ph} \leq 4$) and neutral hydrogen column densities
($10^{20} \leq N_{H} \leq 10^{22}$). 
Sources with $S/N>1.5$ ($open$ $circles$) and
sources with $S/N>2.0$ ($red$ $closed$ $circles$) are plotted.
Most sources have absorption in the range
$10^{20} \lesssim N_{H} \lesssim 10^{21}~cm^{-2}$
and the photon index, $1\lesssim\Gamma_{ph}\lesssim2.5$.
We note that the absorbed sources ($N_{H}>10^{21}~cm^{-2}$) in the ChaMP sample are
not statistically significant ($S/N<1.5$ in at least one energy band)
and so are not plotted in Figure 20 and 21.

In Table \ref{tbl-summary}, we summarize the statistical properties of 
the ChaMP X-ray point sources:
number of sources, minimum, maximum, median, and mean for the source counts,
source fluxes, effective exposure times, off axis angle, hardness ratio HR, 
color C21, and color C32.
We define the properties of typical ChaMP X-ray point source as the median values 
of these quantities for sources with $S/N>2.0$.
The typical source fluxes are 
$4.3\times 10^{-15}$ $erg~cm^{-2}~sec^{-1}$ (0.5-2 keV) and 
$11.1\times 10^{-15}$ $erg~cm^{-2}~sec^{-1}$ (2-8 keV), respectively.
The typical flux ranges are $3.7\times 10^{-16}$ $\sim$ $2.5 \times 10^{-11}$
$erg~cm^{-2}~sec^{-1}$ (0.5-2 keV) and
$1.7\times 10^{-15}$ $\sim$ $6.7 \times 10^{-11}$ 
$erg~cm^{-2}~sec^{-1}$ (2-8 keV), respectively.
The flux ranges of ChaMP X-ray sources cover the flux gap between the
$Chandra$ Deep Fields and previous surveys such as $ASCA$ and $ROSAT$,
and fully cover the flux range around the break in the X-ray
number counts \citep{kim06a}. 

Thanks to the ChaMP's medium depth ($9.4\times 10^{-16}$
$\sim$ $5.9 \times 10^{-11}$ $erg~cm^{-2}~sec^{-1}$ in the 0.5-8 keV),  
wide sky coverage area ($\sim 10$ $deg^{2}$), and large number of sources 
($\sim 6,800$), we can investigate
populations and evolution models of cosmic X-ray sources
with small statistical errors.
The ChaMP is a serendipitous survey, therefore it is suitable for investigating 
the field-to-field variations of X-ray sources.
In later ChaMP papers, we will provide number counts
of X-ray point sources and their contributions to the CXRB, spatial angular 
correlations, X-ray color-color analysis, and optical/IR/radio properties 
of X-ray sources.
Also, we will extend our study of X-ray galaxies \citep{kim06c}
and X-ray galaxy clusters \citep{bar06} in ChaMP fields. 

\section{Summary and Conclusions}
1. We present the full ChaMP X-ray point source catalog. The main
catalog contains $6,512$ sources from 130 ChaMP fields and the
supplementary catalog contains $853$ sources from 19 ChaMP fields
which are partly overlapping in their field of view with those in
the main catalog. After eliminating duplications, our catalogs
contain $\sim6,800$ individual point sources, in a sky area of
$\sim10~deg^{2}$.

2. The ChaMP X-ray sources are uniformly reduced with the ChaMP XPIPE
pipeline and carefully confirmed by visual inspections. 
Photometry in eight X-ray energy bands, hardness ratio, and X-ray
colors of ChaMP X-ray point sources are provided. To calculate the
flux of ChaMP X-ray point sources, we also provide the energy
conversion factors ($ECF$s) for each CCD chip and observation for photon
indices, $\Gamma_{ph}=1.2$, 1.4, and 1.7 and Galactic absorption,
$N_H$ and the vignetting corrected effective exposure times.   

3. To understand the sensitivity and reliability of ChaMP X-ray point
sources, we have performed extensive simulations. The detection
probability of ChaMP sources is greater than $95\%$ for source
counts of $\gtrsim30$ and off axis angle of $<5\arcmin$. 
The count recovery rate is $96\pm1\%$
regardless of source counts.  $wavdetect$ tends to underestimate the
net counts for faint sources ($87\%$ for $\lesssim50$ counts). The
false source detection probability is $\sim 1\%$ of the total detected
sources and $\sim80\%$ of these have source counts of $\lesssim30$.

4. Empirical equations for the positional uncertainties were derived
from the ChaMP simulations. The positional uncertainty in ChaMP fields
exponentially increases with off axis angle and
decreases as the source counts increase with a power law form. 
Background fluctuations do not affect the positional
uncertainty in our ChaMP sample. The $68\%$, $90\%$, and $95\%$
confidence levels of equations for the positional uncertainties are
provided.

5. The absolute positional accuracy of the ChaMP X-ray sources is 
$0.7\pm0.4\arcsec$, estimated by matching with $SDSS$ optical sources.

6. The typical ChaMP X-ray point source
in the 0.5-2 keV band
has counts of 24.3, flux
of $4.3\times 10^{-15}$ $erg~cm^{-2}~sec^{-1}$,
effective exposure time of $37.6~ksec$
and off axis angle of $6.3\arcmin$.
In the 2-8 keV band, the typical source
has counts of 18.9, flux
of $11.1\times10^{-15}$ $erg~cm^{-2}~sec^{-1}$,
effective exposure time of $41.7~ksec$
and off axis angle of $6.24\arcmin$.

7. The hardness ratio and X-ray colors were calculated
with a Bayesian approach which models the detected counts as 
a Poisson distribution.
The typical hardness ratio of the
ChaMP X-ray source is $-0.35$, and X-ray colors C21 and C32
are $-0.29$ and $0.36$, respectively.

8. The flux levels (in $erg~cm^{-2}~sec^{-1}$) of sources are
$9.4\times 10^{-16}$ $\sim$ $5.9 \times 10^{-11}$ (0.5-8 keV),
$3.7\times 10^{-16}$ $\sim$ $2.5 \times 10^{-11}$ (0.5-2 keV),
and $1.7\times 10^{-15}$ $\sim$ $6.7 \times 10^{-11}$ (2-8 keV), respectively.
In the X-ray color-color diagram, typical ChaMP sources are located
within absorption range: 
$10^{20} \lesssim N_{H} \lesssim 10^{21}~cm^{-2}$
and within photon index range: $1\lesssim\Gamma_{ph}\lesssim2.5$.

We gratefully acknowledge support for this project under NASA CXC
archival research grant AR4-5017X and AR6-7020X .  PJG, DWK,
HT, and BJW also acknowledge support through NASA Contract 
NAS8-03060 (CXC).
MGL is in part supported by the KOSEF grant (R01-2004-000-10490-0).

{\it Facility:} \facility{CXO (ACIS)}

\clearpage

\begin{center}
\begin{deluxetable}{rclcccrrcrcc}
\tablecaption{List of ChaMP Fields\label{tbl-champ-list}}
\tablewidth{0pt}
\tabletypesize{\tiny}
\tablecolumns{12}\tablehead{
\colhead{Obs.}&
\colhead{Sequence}&
\colhead{Target}&
\colhead{R.A.}&
\colhead{Decl.}&
\colhead{$N_{H}$}&
\colhead{Obs.}&
\colhead{Exposure}&
\colhead{Aim Point}&
\colhead{CCD}&
\colhead{Multi}&
\colhead{Paper I}\\
\colhead{ID}&
\colhead{Number}&
\colhead{Name}&
\colhead{(J2000)}&
\colhead{(J2000)}&
\colhead{$(10^{20}cm^{-2})$}&
\colhead{Date}&
\colhead{($ksec$)}&
\colhead{Detector}&
\colhead{Used}&
\colhead{ID}&
\colhead{Used}\\
\colhead{(1)}&
\colhead{(2)}&
\colhead{(3)}&
\colhead{(4)}&
\colhead{(5)}&
\colhead{(6)}&
\colhead{(7)}&
\colhead{(8)}&
\colhead{(9)}&
\colhead{(10)}&
\colhead{(11)}&
\colhead{(12)} 
}
\startdata
\dataset[ADS/Sa.CXO#obs/520]{520} & 800028 & MS 0015.9+1609       & $ 00:18:32.7 $ & $ +16:30:04.0 $ &  4.06 &  2000 Aug 18 &   60.99 & ACIS-I &     01236 &    &  Y \\ 
\dataset[ADS/Sa.CXO#obs/1595]{1595} & 700211 & 3C9                  & $ 00:20:24.0 $ & $ +15:42:09.1 $ &  4.16 &  2001 Jun 10 &   13.27 & ACIS-S &    235678 &    &    \\ 
\dataset[ADS/Sa.CXO#obs/2098]{2098} & 700288 & LBQS 0019+0107       & $ 00:22:28.0 $ & $ +01:23:37.6 $ &  2.92 &  2001 Dec 08 &    4.70 & ACIS-S &    235678 &    &    \\ 
\dataset[ADS/Sa.CXO#obs/2099]{2099} & 700289 & LBQS 0021-0213       & $ 00:24:10.6 $ & $ -01:56:12.1 $ &  2.95 &  2001 Aug 20 &    6.65 & ACIS-S &    235678 &    &    \\ 
\dataset[ADS/Sa.CXO#obs/929]{929} & 800105 & CL 0024+17           & $ 00:26:31.8 $ & $ +17:10:57.4 $ &  4.19 &  2000 Sep 06 &   36.72 & ACIS-S &    245678 &    &    \\ 
\dataset[ADS/Sa.CXO#obs/2100]{2100} & 700290 & LBQS 0025-0151       & $ 00:27:34.6 $ & $ -01:35:25.9 $ &  2.81 &  2001 Oct 31 &    4.40 & ACIS-S &    235678 &    &    \\ 
\dataset[ADS/Sa.CXO#obs/2101]{2101} & 700291 & LBQS 0029+0017       & $ 00:31:35.5 $ & $ +00:34:58.7 $ &  2.41 &  2001 Jun 20 &    3.82 & ACIS-S &    235678 &    &    \\ 
\dataset[ADS/Sa.CXO#obs/2242]{2242} & 900069 & GSGP4X:048           & $ 00:57:17.9 $ & $ -27:22:23.8 $ &  1.69 &  2000 Dec 18 &    6.66 & ACIS-S &    235678 & 12 &  Y \\
\enddata
\tablecomments{
Col. (1): observational ID.
Col. (2): sequence number. 
Col. (3): target name. 
Col. (4) and (5): the right ascension in units of hours, minutes, and seconds,
and declination in units of degrees, arcminutes, and arcseconds
of the aim point in J2000.
Col. (6): the galactic extinction in units of
$10^{20}~cm^{-2}$ from \cite{sta92}.
Col. (7): the date of observation. 
Col. (8): the net exposure time estimated after removing background flares
within the aim pointed CCD chip. 
Col. (9): aim point detector type.
Col. (10): used CCD chips (CCD ID=0-3 for ACIS-I and 4-9 for ACIS-S). 
Col. (11): multiple observation ID indicates overlapping fields, 
labeled with the same ID.
Col. (12): 62 ChaMP fields included in Paper I are marked as Y.}
\end{deluxetable}
\end{center}
\clearpage

\clearpage
\begin{center}
\begin{deluxetable}{lr}
\tablecaption{Definition of Energy Bands and X-ray Colors\label{tbl-ebands}}
\tablewidth{0pt}
\tabletypesize{\normalsize}
\tablecolumns{2}\tablehead{
\colhead{Band}&
\colhead{Definition}
}
\startdata
Broad (B)       &    $0.3-8$ keV \\
Soft (S)        &    $0.3-2.5$ keV \\
Hard (H)        &    $2.5-8$ keV \\
Soft1 (S$_{1}$) &    $0.3-0.9$ keV \\
Soft2 (S$_{2}$) &    $0.9-2.5$ keV \\
\hline
Conventional Broad (Bc)       &    $0.5-8$ keV \\
Conventional Soft (Sc)        &    $0.5-2$ keV \\
Conventional Hard (Hc)        &    $2-8$ keV \\
\hline
Hardness Ratio $HR$  &    (Hc$-$Sc)$/$(Hc$+$Sc) \\
X-ray Color $C21$          &    $-log$(S$_{2}$)$+log$(S$_{1}$)=$log$(S$_{1}$/S$_{2}$) \\
X-ray Color $C32$          &    $-log$(H)$+log$(S$_{2}$)=$log$(S$_{2}$/H) \\
\enddata
\end{deluxetable}
\end{center}

\clearpage
\begin{center}
\begin{deluxetable}{lc|rrrrrr|rrrrrr|rrrrrr}
\rotate
\tablecaption{Energy Conversion Factor\label{tbl-energy-con-factor}}
\tablewidth{0pt}
\tabletypesize{\tiny}
\tablecolumns{20}\tablehead{

\colhead{Obs.}&
\colhead{CCD}&
\multicolumn{6}{c}{$\Gamma_{ph}=1.2$} &
\multicolumn{6}{c}{$\Gamma_{ph}=1.4$} &
\multicolumn{6}{c}{$\Gamma_{ph}=1.7$} \\
\colhead{ID}&
\colhead{ID}&
\colhead{B}&
\colhead{S}&
\colhead{H}&
\colhead{Bc}&
\colhead{Sc}&
\colhead{Hc}&
\colhead{B}&
\colhead{S}&
\colhead{H}&
\colhead{Bc}&
\colhead{Sc}&
\colhead{Hc}&
\colhead{B}&
\colhead{S}&
\colhead{H}&
\colhead{Bc}&
\colhead{Sc}&
\colhead{Hc}\\
\colhead{(1)}&
\colhead{(2)}&
\colhead{(3)}&
\colhead{(4)}&
\colhead{(5)}&
\colhead{(6)}&
\colhead{(7)}&
\colhead{(8)}&
\colhead{(9)}&
\colhead{(10)}&
\colhead{(11)}&
\colhead{(12)}&
\colhead{(13)}&
\colhead{(14)}&
\colhead{(15)}&
\colhead{(16)}&
\colhead{(17)}&
\colhead{(18)}&
\colhead{(19)}&
\colhead{(20)}
}
\startdata
  135 &     2 &     130.25 &      64.24 &     289.26 &     128.03 &      50.19 &     262.14 &     117.76 &      65.36 &     277.79 &     114.15 &      50.74 &     250.33 &     104.04 &      68.04 &     262.06 &      97.53 &      51.74 &     234.34 \\
  135 &     3 &     131.44 &      65.01 &     289.92 &     129.12 &      50.77 &     263.01 &     118.98 &      66.21 &     278.41 &     115.23 &      51.36 &     251.20 &     105.31 &      69.03 &     262.70 &      98.60 &      52.44 &     235.21 \\
  135 &     5 &     103.05 &      44.69 &     341.86 &     108.74 &      37.58 &     289.69 &      87.98 &      43.67 &     324.75 &      93.20 &      37.10 &     272.63 &      71.00 &      42.50 &     301.68 &      75.06 &      36.43 &     249.91 \\
  135 &     6 &     115.00 &      56.38 &     259.13 &     113.61 &      44.20 &     235.68 &     103.57 &      57.09 &     249.59 &     101.04 &      44.53 &     225.69 &      90.79 &      58.94 &     236.46 &      85.88 &      45.16 &     212.07 \\
  135 &     7 &      85.12 &      38.20 &     248.29 &      89.43 &      32.10 &     217.62 &      73.37 &      37.40 &     237.38 &      77.30 &      31.73 &     206.28 &      59.93 &      36.52 &     222.43 &      62.89 &      31.19 &     190.93 \\
  325 &     2 &     135.69 &      65.77 &     314.52 &     133.17 &      51.72 &     277.16 &     122.42 &      67.05 &     301.06 &     118.41 &      52.31 &     263.12 &     107.98 &      70.02 &     282.68 &     100.88 &      53.40 &     244.09 \\
  325 &     3 &     136.58 &      66.61 &     311.79 &     133.90 &      52.20 &     276.84 &     123.44 &      67.96 &     298.62 &     119.24 &      52.85 &     263.16 &     109.16 &      71.07 &     280.62 &     101.81 &      54.01 &     244.65 \\
\enddata
\tablecomments{
Col. (1): observation ID. 
Col. (2): CCD ID.
Col. (3) - Col. (8): energy conversion factors in units of $erg$ $cm^{-2}$
$count^{-1}$ assumimg a photon index of
$\Gamma_{ph}=1.2$ in the B, S, H,
Bc, Sc, and Hc bands, respectively.
Col. (9) - Col. (14): energy conversion factors in units of $erg$ $cm^{-2}$
$count^{-1}$ assuming a photon index 
$\Gamma_{ph}=1.4$ in the B, S, H,
Bc, Sc, and Hc bands, respectively.
Col. (15) - Col. (20): energy conversion factors in units of $erg$ $cm^{-2}$
$count^{-1}$ assuming a photon index
of $\Gamma_{ph}=1.7$ in the B, S, H,
Bc, Sc, and Hc bands, respectively.
The complete version of this table is in the electronic edition of
the $Astrophysical~Journal~Supplement$.  The printed edition contains only a sample.
}
\end{deluxetable}
\end{center}
\clearpage

\clearpage
\begin{center}
\begin{deluxetable}{cl}
\tablecaption{Definition of Source Flag \label{tbl-flag}}
\tablewidth{0pt}
\tabletypesize{\normalsize}
\tablecolumns{2}\tablehead{\colhead{Flag}&
\colhead{Definition}}
\startdata
1n & False X-ray Sources \\
11 & False source due to a hot pixel or by a bad bias value \\
12 & False source due to a bad column \\
13 & False source along the readout direction of a very strong source \\
14 & False source due to the FEP 0/3 problem \\
15 & Double sources detected due to the PSF effect  \\
\hline
2n & It is not certain whether it is a valid X-ray source \\
21 & V\&Ver found it might be a spurious source \\
\hline
3n & Source properties may be subject to a large uncertainty \\
31 & Bad pixel/column exists within source extraction radius \\
37 & Pile-up (see $Chandra$ POG 2004) \\
38 & Uncertain source position by flag=015 \\
\hline
5n & Other cases \\
51 & Source is extended \\
52 & Same source candidate in multiple observations \\
53 & Target of observation \\
54 & X-ray jet \\
55 & Variable source (see Paper I) \\
\hline
6n & Flagged by the automatic pipeline $xapphot$ \\
61 & Source region falls near the edge of the chip \\
62 & Small portion of source region overlaps with nearby point source \\
63 & Large portion of source region overlaps with nearby point source \\
64 & Small portion of source region overlaps with nearby extended source \\
65 & Large portion of source region overlaps with nearby extended source \\
66 & Background region overlaps with a nearby point source \\
67 & Background region overlaps with a nearby extended source \\
68 & Source region falls inside the extended source \\
\enddata
\end{deluxetable}
\end{center}

\clearpage
\begin{center}
\begin{deluxetable}{ccccccrrrrc}
\tablecaption{The Main ChaMP X-ray Point Sources\label{tbl-main-list}}
\tablewidth{0pt}
\tabletypesize{\scriptsize}
\tablecolumns{11}\tablehead{
\colhead{}&
\colhead{Obs.}&
\colhead{CCD}&
\colhead{Source}&
\colhead{R.A.}&
\colhead{Decl.}&
\colhead{PU}&
\colhead{$D_{off-axis}$}&
\colhead{Radius}&
\colhead{Exposure}&
\colhead{}\\
\colhead{Source Name}&
\colhead{ID}&
\colhead{ID}&
\colhead{No.}&
\colhead{(deg)}&
\colhead{(deg)}&
\colhead{(arcsec)}&
\colhead{(arcmin)}&
\colhead{(arcsec)}&
\colhead{(ksec)}&
\colhead{Flag}\\
\colhead{(1)}&
\colhead{(2)}&
\colhead{(3)}&
\colhead{(4)}&
\colhead{(5)}&
\colhead{(6)}&
\colhead{(7)}&
\colhead{(8)}&
\colhead{(9)}&
\colhead{(10)}&
\colhead{(11)}}
\startdata
CXOMP J001709.7+162604 &   520 &   6 &     1 &        4.290515 &       16.434570 &        60: &      20.28 &      40.00 &      51.72 &  66                                       \\
CXOMP J001716.0+162722 &   520 &   6 &     4 &        4.316865 &       16.456352 &        60: &      18.56 &      40.00 &      53.94 &  66                                       \\
CXOMP J001739.7+162330 &   520 &   6 &     2 &        4.415671 &       16.391680 &       3.12 &      14.28 &      27.02 &      59.54 &                                           \\
CXOMP J001743.5+163144 &   520 &   2 &    24 &        4.431394 &       16.528941 &       7.21 &      11.90 &      18.98 &      48.00 &                                           \\
CXOMP J001756.6+163006 &   520 &   2 &    23 &        4.486132 &       16.501804 &       2.44 &       8.63 &      10.69 &      55.58 &                                           \\
CXOMP J001758.4+162647 &   520 &   2 &    21 &        4.493674 &       16.446530 &       2.79 &       8.82 &      11.11 &      58.17 &  55                                       \\
CXOMP J001758.9+163119 &   520 &   2 &    20 &        4.495359 &       16.521831 &       1.79 &       8.20 &       9.75 &      58.94 &  66                                       \\
CXOMP J001801.0+163139 &   520 &   2 &    19 &        4.504432 &       16.527670 &       2.14 &       7.75 &       8.77 &      54.89 &  66                                       \\
\enddata
\tablecomments{
Col. (1): ChaMP source name given by right ascension
in units of hours, minutes, and seconds,
and declination in units of degrees, arcminutes, and arcseconds in J2000. 
Col. (2): observation ID. 
Col. (3): CCD ID (0-3 for ACIS-I and 4-9 for ACIS-S observation).
Col. (4): source number assigned by $wavdetect$ as 
marked in the public X-ray image of the ChaMP web site
(http://hea-www.harvard.edu/CHAMP).
Col. (5) and (6):  right ascension and declination in units of degrees
after refinement
and aspect correction. 
Col. (7): positional uncertainty in units of arcseconds estimated by
the $95\%$ confidence level of the empirical formula
(see equation (12) in \S 4.2). 
Col. (8): off axis angle from the aim point of the observation
in units of arcminutes. 
Col. (9): source extraction radius in units of arcseconds (The $95\%$ encircled
energy radius at 1.5 keV. See \S 3.2.1). 
Col. (10): mean effective exposure time in units of ksec after 
vignetting correction (see \S 3.2.4).
Col. (11): source flag (see Table \ref{tbl-flag}). 
The complete version of this table is in the electronic edition of
the $Astrophysical~Journal~Supplement$.  The printed edition contains only a sample.
}
\end{deluxetable}
\end{center}
\clearpage

\clearpage
\begin{center}
\begin{deluxetable}{ccccccrrrrc}
\tablecaption{The Supplementary ChaMP X-ray Point Sources \label{tbl-sub-list}}
\tablewidth{0pt}
\tabletypesize{\scriptsize}
\tablecolumns{11}\tablehead{
\colhead{}&
\colhead{Obs.}&
\colhead{CCD}&
\colhead{Source}&
\colhead{R.A.}&
\colhead{Decl.}&
\colhead{PU}&
\colhead{$D_{off-axis}$}&
\colhead{Radius}&
\colhead{Exposure}&
\colhead{}\\
\colhead{Source Name}&
\colhead{ID}&
\colhead{ID}&
\colhead{No.}&
\colhead{(deg)}&
\colhead{(deg)}&
\colhead{(arcsec)}&
\colhead{(arcmin)}&
\colhead{(arcsec)}&
\colhead{(ksec)}&
\colhead{Flag}\\
\colhead{(1)}&
\colhead{(2)}&
\colhead{(3)}&
\colhead{(4)}&
\colhead{(5)}&
\colhead{(6)}&
\colhead{(7)}&
\colhead{(8)}&
\colhead{(9)}&
\colhead{(10)}&
\colhead{(11)}}
\startdata
CXOMP J005618.9-272429 &  2242 &   3 &     4 &       14.079076 &      -27.408259 &      15.58 &      13.24 &      23.32 &       8.60 &  66                                       \\
CXOMP J005622.4-272223 &  2242 &   3 &     2 &       14.093514 &      -27.373174 &       6.50 &      12.31 &      20.13 &       8.33 &                                           \\
CXOMP J005622.8-272746 &  2242 &   3 &     3 &       14.095076 &      -27.462953 &      13.97 &      13.35 &      23.71 &       8.35 &                                           \\
CXOMP J005630.6-273141 &  2242 &   2 &     4 &       14.127555 &      -27.528123 &      12.27 &      14.01 &      26.06 &       9.67 &  66                                       \\
CXOMP J005633.8-272954 &  2242 &   2 &     2 &       14.140895 &      -27.498360 &       5.52 &      12.33 &      20.18 &      10.15 &  66                                       \\
CXOMP J005644.3-274936 &  2243 &   3 &     5 &       14.184665 &      -27.826692 &      18.16 &      11.18 &      16.97 &       8.13 &                                           \\
CXOMP J005645.6-272444 &  2242 &   3 &     1 &       14.190344 &      -27.412292 &       3.66 &       7.52 &       8.29 &       9.59 &                                           \\
CXOMP J005645.9-275522 &  2243 &   3 &     4 &       14.191467 &      -27.922784 &       8.72 &      14.68 &      28.46 &       7.53 &                                           \\
\enddata
\tablecomments{
Col. (1): ChaMP source name given by right ascension
in units of hours, minutes, and seconds,
and declination in units of degrees, arcminutes, and arcseconds in J2000.
Col. (2): observation ID.
Col. (3): CCD ID (0-3 for ACIS-I and 4-9 for ACIS-S observation).
Col. (4): source number assigned by $wavdetect$ 
as marked in the public X-ray image of the ChaMP web site
(http://hea-www.harvard.edu/CHAMP).
Col. (5) and (6): right ascension and declination in units of degrees
after refinement
and aspect correction.
Col. (7): positional uncertainty in units of arcseconds estimated by
the $95\%$ confidence level of the empirical formula
(see equation (12) in \S 4.2).
Col. (8): off axis angle from the aim point of the observation
in units of arcminutes.
Col. (9): source extraction radius in units of arcseconds (The 95\% energy encircled
radius at 1.5 keV. See \S 3.2.1).
Col. (10): mean effective exposure time in units of ksec after
vignetting correction (see \S 3.2.4).
Col. (11): source flag (see Table \ref{tbl-flag}).
The complete version of this table is in the electronic edition of
the $Astrophysical~Journal~Supplement$.  The printed edition contains only a sample.
}
\end{deluxetable}
\end{center}
\clearpage

\clearpage
\begin{center}
\begin{deluxetable}{cccccrcccc}
\tablecaption{The Multiply-Observed ChaMP X-ray Point Source Candidates\label{tbl-multi-list}}
\tablewidth{0pt}
\tabletypesize{\scriptsize}
\tablecolumns{11}\tablehead{
\colhead{}&
\colhead{Multi}&
\colhead{Obs.}&
\colhead{CCD}&
\colhead{Source}&
\colhead{Obs.}&
\colhead{}&
\colhead{}&
\colhead{$D_{off-axis}$}&
\colhead{PU}\\
\colhead{Source Name}&
\colhead{ID}&
\colhead{ID}&
\colhead{ID}&
\colhead{No.}&
\colhead{Date}&
\colhead{netB}&
\colhead{rateB}&
\colhead{(arcmin)}&
\colhead{(arcsec)}\\
\colhead{(1)}&
\colhead{(2)}&
\colhead{(3)}&
\colhead{(4)}&
\colhead{(5)}&
\colhead{(6)}&
\colhead{(7)}&
\colhead{(8)}&
\colhead{(9)}&
\colhead{(10)}
}
\startdata
CXOMP J025656.1-331757 &    1 &   135 &     6 &     1 & Feb 15 2000 & $       3.00\pm      2.94 $ & $       4.20\pm      4.11 $ &  $       2.65 $ & $       1.84 $  \\
CXOMP J025656.1-331756 &    1 &   815 &     6 &     2 & Jan  2 2000 & $       2.00\pm      2.66 $ & $       1.30\pm      1.73 $ &  $       2.43 $ & $       2.13 $  \\
CXOMP J025655.6-332109 &    2 &   135 &     6 &     2 & Feb 15 2000 & $       5.86\pm      3.60 $ & $       6.65\pm      4.09 $ &  $       5.40 $ & $       2.74 $  \\
CXOMP J025655.9-332106 &    2 &   815 &     6 &     4 & Jan  2 2000 & $       3.62\pm      3.18 $ & $       2.47\pm      2.17 $ &  $       5.31 $ & $       4.66 $  \\
CXOMP J025647.8-331524 &    3 &   135 &     7 &     1 & Feb 15 2000 & $      24.81\pm      6.08 $ & $      27.44\pm      6.72 $ &  $       0.65 $ & $       0.38 $  \\
CXOMP J025647.8-331524 &    3 &   815 &     7 &     1 & Jan  2 2000 & $      17.85\pm      5.33 $ & $      15.97\pm      4.77 $ &  $       0.65 $ & $       0.46 $  \\
CXOMP J025642.6-331521 &    4 &   135 &     7 &     2 & Feb 15 2000 & $      24.00\pm      5.98 $ & $      26.69\pm      6.65 $ &  $       1.23 $ & $       0.45 $  \\
CXOMP J025642.5-331520 &    4 &   815 &     7 &     2 & Jan  2 2000 & $      27.81\pm      6.36 $ & $      25.27\pm      5.78 $ &  $       1.52 $ & $       0.46 $  \\
\enddata
\tablecomments{
Col. (1): ChaMP source name given by right ascension
in units of hours, minutes, and seconds,
and. declination in units of degrees, arcminutes, and arcseconds in J2000. 
Col. (2): multiple observation ID indicates the same source candidate in overlapping fields,
labeled with the same ID.
Col. (3): observation ID. 
Col. (4): CCD ID (0-3 for ACIS-I and 4-9 for ACIS-S).
Col. (5): source number assigned by $wavdetect$ as marked in
the public X-ray images of the ChaMP web site (http://hea-www.harvard.edu/CHAMP). 
Col. (6): date of observation. 
Col. (7): net counts in the 0.3-8 keV band and errors. 
Col. (8): count rates in the 0.3-8 keV band and errors. 
Col. (9): off axis angle from the aim point of the observation
in units of arcminutes. 
Col. (10): positional uncertainty in units of arcseconds estimated by
the $95\%$ confidence level of the empirical formula
(see equation (12) in \S 4.2). 
The errors in net counts are given following Gehrels
statistics \citep{geh86}. 
The complete version of this table is in the electronic edition of
the $Astrophysical~Journal~Supplement$. The printed edition contains only a sample.
}
\end{deluxetable}
\end{center}
\clearpage

\clearpage
\begin{center}
\begin{deluxetable}{lcccccccc}
\tablecaption{The Main ChaMP X-ray Point Source Photometry \label{tbl-main-phot}}
\tablewidth{0pt}
\tabletypesize{\tiny}
\tablecolumns{9}\tablehead{
\colhead{Source Name}&
\colhead{B}&
\colhead{S$_{1}$}&
\colhead{S$_{2}$}&
\colhead{S}&
\colhead{H}&
\colhead{Bc}&
\colhead{Sc}&
\colhead{Hc}\\
\colhead{(1)}&
\colhead{(2)}&
\colhead{(3)}&
\colhead{(4)}&
\colhead{(5)}&
\colhead{(6)}&
\colhead{(7)}&
\colhead{(8)}&
\colhead{(9)}}
\startdata
CXOMP J001709.7+162604 & $      10.20\pm      17.71 $ & $       5.55\pm       9.03 $ & $      14.43\pm      10.48 $ & $      19.98\pm      13.40 $ & $     $-----$ $ & $       6.13\pm      17.38 $ & $      17.75\pm      11.51 $ & $     $-----$ $ \\
CXOMP J001716.0+162722 & $      94.40\pm      20.17 $ & $      25.10\pm       9.91 $ & $      48.11\pm      12.09 $ & $      73.21\pm      15.20 $ & $      21.19\pm      13.91 $ & $      90.00\pm      20.10 $ & $      60.42\pm      13.42 $ & $      29.58\pm      14.97 $ \\
CXOMP J001739.7+162330 & $     233.49\pm      20.14 $ & $      51.89\pm       9.68 $ & $     131.11\pm      14.06 $ & $     183.00\pm      16.66 $ & $      50.48\pm      12.03 $ & $     234.00\pm      20.48 $ & $     169.84\pm      15.74 $ & $      64.16\pm      13.11 $ \\
CXOMP J001743.5+163144 & $      20.29\pm       9.20 $ & $     $-----$ $ & $      16.27\pm       6.38 $ & $      15.86\pm       7.24 $ & $       4.43\pm       6.39 $ & $      21.71\pm       9.45 $ & $      18.72\pm       6.76 $ & $       3.00\pm       6.61 $ \\
CXOMP J001756.6+163006 & $      34.54\pm       8.17 $ & $       6.26\pm       4.14 $ & $      19.36\pm       6.00 $ & $      25.62\pm       6.84 $ & $       8.92\pm       5.26 $ & $      30.65\pm       8.28 $ & $      20.31\pm       6.10 $ & $      10.34\pm       5.60 $ \\
CXOMP J001758.4+162647 & $      29.16\pm       8.14 $ & $       4.83\pm       3.99 $ & $      14.49\pm       5.59 $ & $      19.32\pm       6.41 $ & $       9.85\pm       5.74 $ & $      28.73\pm       8.53 $ & $      20.11\pm       6.21 $ & $       8.63\pm       5.85 $ \\
CXOMP J001758.9+163119 & $      58.38\pm       9.57 $ & $       4.80\pm       3.99 $ & $      41.65\pm       7.78 $ & $      46.45\pm       8.36 $ & $      11.93\pm       5.48 $ & $      56.72\pm       9.83 $ & $      43.87\pm       8.00 $ & $      12.85\pm       5.71 $ \\
\enddata
\tablecomments{
Col. (1): ChaMP source name given by right ascension
in units of hours, minutes, and seconds,
and declination in units of degrees, arcminutes, and arcseconds in J2000. 
Col. (2): net counts and errors in the B ($0.3-8$ keV) band. 
Col. (3): net counts and errors in the S$_{1}$ ($0.3-0.9$ keV) band. 
Col. (4): net counts and errors in the S$_{2}$ ($0.9-2.5$ keV) band. 
Col. (5): net counts and errors in the S ($0.3-2.5$ keV) band. 
Col. (6): net counts and errors in the H ($2.5-8$ keV) band. 
Col. (7): net counts and errors in the Bc ($0.5-8$ keV) band. 
Col. (8): net counts and errors in the Sc ($0.5-2$ keV) band. 
Col. (9): net counts and errors in the Hc ($2-8$ keV) band. 
For detail information about the X-ray photometry, see \S 3.2.
The errors of the net counts are given by following Gehrels statistics \citep{geh86}.
The net counts which are not determined are marked as $-$.
The complete version of this table is in the electronic edition of
the $Astrophysical~Journal~Supplement$.  The printed edition contains only a sample.
}
\end{deluxetable}
\end{center}
\clearpage

\clearpage
\begin{center}
\begin{deluxetable}{lcccccccc}
\tablecaption{The Supplementary ChaMP X-ray Point Source Photometry \label{tbl-sub-phot}}
\tablewidth{0pt}
\tabletypesize{\tiny}
\tablecolumns{9}\tablehead{
\colhead{Source Name}&
\colhead{B}&
\colhead{S$_{1}$}&
\colhead{S$_{2}$}&
\colhead{S}&
\colhead{H}&
\colhead{Bc}&
\colhead{Sc}&
\colhead{Hc}\\
\colhead{(1)}&
\colhead{(2)}&
\colhead{(3)}&
\colhead{(4)}&
\colhead{(5)}&
\colhead{(6)}&
\colhead{(7)}&
\colhead{(8)}&
\colhead{(9)}}
\startdata
CXOMP J005618.9-272429 & $      16.06\pm       6.51 $ & $       0.97\pm       2.96 $ & $      12.10\pm       4.99 $ & $      13.07\pm       5.36 $ & $       2.99\pm       4.48 $ & $      13.88\pm       6.71 $ & $       7.72\pm       4.45 $ & $       6.16\pm       5.02 $ \\
CXOMP J005622.4-272223 & $      40.97\pm       8.15 $ & $      12.45\pm       4.85 $ & $      24.58\pm       6.28 $ & $      37.03\pm       7.48 $ & $       3.94\pm       4.16 $ & $      36.65\pm       8.23 $ & $      30.40\pm       6.82 $ & $       6.25\pm       4.61 $ \\
CXOMP J005622.8-272746 & $      14.02\pm       6.43 $ & $       3.71\pm       3.62 $ & $       5.99\pm       4.31 $ & $       9.70\pm       5.14 $ & $       4.32\pm       4.63 $ & $      13.42\pm       6.82 $ & $       7.13\pm       4.61 $ & $       6.29\pm       5.03 $ \\
CXOMP J005630.6-273141 & $      32.02\pm       8.48 $ & $       8.35\pm       4.74 $ & $      18.61\pm       6.11 $ & $      26.96\pm       7.27 $ & $       5.07\pm       5.16 $ & $      27.21\pm       8.52 $ & $      20.34\pm       6.50 $ & $       6.87\pm       5.51 $ \\
CXOMP J005633.8-272954 & $      59.47\pm       9.69 $ & $      12.47\pm       4.99 $ & $      36.54\pm       7.48 $ & $      49.01\pm       8.56 $ & $      10.46\pm       5.37 $ & $      57.77\pm      10.00 $ & $      46.50\pm       8.28 $ & $      11.27\pm       5.60 $ \\
CXOMP J005644.3-274936 & $      14.48\pm       5.91 $ & $       0.94\pm       2.67 $ & $       9.03\pm       4.44 $ & $       9.97\pm       4.73 $ & $       4.51\pm       4.32 $ & $      14.06\pm       6.30 $ & $       9.27\pm       4.44 $ & $       4.79\pm       4.47 $ \\
CXOMP J005645.6-272444 & $      10.36\pm       4.58 $ & $       1.69\pm       2.66 $ & $       5.59\pm       3.60 $ & $       7.28\pm       3.97 $ & $       3.08\pm       3.19 $ & $      10.36\pm       5.09 $ & $       7.33\pm       3.97 $ & $       3.02\pm       3.19 $ \\
\enddata
\tablecomments{
Col. (1): ChaMP source name given by right ascension
in units of hours, minutes, and seconds,
and declination in units of degrees, arcminutes, and arcseconds in J2000.
Col. (2): net counts and errors in the B ($0.3-8$ keV) band.
Col. (3): net counts and errors in the S$_{1}$ ($0.3-0.9$ keV) band.
Col. (4): net counts and errors in the S$_{2}$ ($0.9-2.5$ keV) band.
Col. (5): net counts and errors in the S ($0.3-2.5$ keV) band.
Col. (6): net counts and errors in the H ($2.5-8$ keV) band.
Col. (7): net counts and errors in the Bc ($0.5-8$ keV) band.
Col. (8): net counts and errors in the Sc ($0.5-2$ keV) band.
Col. (9): net counts and errors in the Hc ($2.0-8$ keV) band.
For detail information about the X-ray photometry, see \S 3.2.
The errors of the net counts are given by following Gehrels 
statistics \citep{geh86}.
The net counts which are not determined are marked as $-$.
The complete version of this table is in the electronic edition of
the $Astrophysical~Journal~Supplement$. The printed edition contains only a sample.
}
\end{deluxetable}
\end{center}
\clearpage

\clearpage
\begin{center}
\begin{deluxetable}{crrrrrrrrr}
\tablecaption{The Main ChaMP X-ray Point Source Hardness Ratio and X-ray Colors \label{tbl-main-color}}
\tablewidth{0pt}
\tabletypesize{\scriptsize}
\tablecolumns{10}\tablehead{
\colhead{Source Name}&
\colhead{HR\_l68}&
\colhead{HR\_mean}&
\colhead{HR\_u68}&
\colhead{C21\_l68}&
\colhead{C21\_mode}&
\colhead{C21\_u68}&
\colhead{C32\_l68}&
\colhead{C32\_mode}&
\colhead{C32\_u68}\\
\colhead{(1)}&
\colhead{(2)}&
\colhead{(3)}&
\colhead{(4)}&
\colhead{(5)}&
\colhead{(6)}&
\colhead{(7)}&
\colhead{(8)}&
\colhead{(9)}&
\colhead{(10)}
}
\startdata
CXOMP J001709.7+162604 & $      -1.00 $ & $      -0.46 $ & $      -0.32 $ & $      -0.70 $ & $      -0.22 $ & $       0.20 $ & $      -0.10 $ & $       0.36 $ & $       0.96 $ \\
CXOMP J001716.0+162722 & $      -0.53 $ & $      -0.35 $ & $      -0.10 $ & $      -0.45 $ & $      -0.28 $ & $      -0.08 $ & $       0.02 $ & $       0.23 $ & $       0.53 $ \\
CXOMP J001739.7+162330 & $      -0.53 $ & $      -0.45 $ & $      -0.37 $ & $      -0.48 $ & $      -0.39 $ & $      -0.32 $ & $       0.30 $ & $       0.37 $ & $       0.51 $ \\
CXOMP J001743.5+163144 & $      -0.90 $ & $      -0.54 $ & $      -0.33 $ & $      -1.22 $ & $      -0.70 $ & $      -0.36 $ & $       0.04 $ & $       0.34 $ & $       0.74 $ \\
CXOMP J001756.6+163006 & $      -0.53 $ & $      -0.31 $ & $      -0.10 $ & $      -0.66 $ & $      -0.44 $ & $      -0.22 $ & $       0.08 $ & $       0.28 $ & $       0.52 $ \\
CXOMP J001758.4+162647 & $      -0.62 $ & $      -0.38 $ & $      -0.14 $ & $      -0.68 $ & $      -0.40 $ & $      -0.14 $ & $      -0.08 $ & $       0.14 $ & $       0.38 $ \\
CXOMP J001758.9+163119 & $      -0.67 $ & $      -0.53 $ & $      -0.40 $ & $      -1.10 $ & $      -0.84 $ & $      -0.62 $ & $       0.35 $ & $       0.50 $ & $       0.68 $ \\
CXOMP J001801.0+163139 & $      -0.36 $ & $      -0.15 $ & $       0.08 $ & $      -0.76 $ & $      -0.44 $ & $      -0.20 $ & $      -0.14 $ & $       0.05 $ & $       0.26 $ \\
\enddata
\tablecomments{
Col. (1): ChaMP source name given by right ascension
in units of hours, minutes, and seconds,
and declination in units of degrees, arcminutes, and arcseconds in J2000. 
Col. (2): the $68\%$ lower limit of hardness ratio HR. 
Col. (3): the mean of hardness ratio (HR$=$(Hc$-$Sc)$/$(Hc$+$Sc)). 
Col. (4): the $68\%$ upper limit of hardness ratio HR. 
Col. (5): the $68\%$ lower limit of color C21. 
Col. (6): the mode of X-ray color C21 ($=log$(S$_{1}/$S$_{2})$). 
Col. (7): the $68\%$ upper limit of color C21. 
Col. (8): the $68\%$ lower limit of color C32. 
Col. (9): the mode of X-ray color C32 ($=log$(S$_{2}/$H)).
Col. (10): the $68\%$ upper limit of color C32. 
For detail information about photometry and color determinations, see \S 3.2.
See Table \ref{tbl-ebands} for definition of energy bands.
The complete version of this table is in the electronic edition of
the $Astrophysical~Journal~Supplement$. The printed edition contains only a sample.
}
\end{deluxetable}
\end{center}
\clearpage

\clearpage
\begin{center}
\begin{deluxetable}{crrrrrrrrr}
\tablecaption{The Supplementary ChaMP X-ray Point Source Hardness Ratio and X-ray Colors \label{tbl-sub-color}}
\tablewidth{0pt}
\tabletypesize{\scriptsize}
\tablecolumns{10}\tablehead{
\colhead{Source Name}&
\colhead{HR\_l68}&
\colhead{HR\_mean}&
\colhead{HR\_u68}&
\colhead{C21\_l68}&
\colhead{C21\_mode}&
\colhead{C21\_u68}&
\colhead{C32\_l68}&
\colhead{C32\_mode}&
\colhead{C32\_u68}\\
\colhead{(1)}&
\colhead{(2)}&
\colhead{(3)}&
\colhead{(4)}&
\colhead{(5)}&
\colhead{(6)}&
\colhead{(7)}&
\colhead{(8)}&
\colhead{(9)}&
\colhead{(10)}
}
\startdata
CXOMP J005618.9-272429 & $      -0.43 $ & $      -0.11 $ & $       0.26 $ & $      -1.14 $ & $      -0.70 $ & $      -0.36 $ & $       0.10 $ & $       0.40 $ & $       0.80 $ \\
CXOMP J005622.4-272223 & $      -0.80 $ & $      -0.63 $ & $      -0.48 $ & $      -0.44 $ & $      -0.28 $ & $      -0.13 $ & $       0.40 $ & $       0.64 $ & $       0.98 $ \\
CXOMP J005622.8-272746 & $      -0.40 $ & $      -0.06 $ & $       0.32 $ & $      -0.52 $ & $      -0.16 $ & $       0.16 $ & $      -0.28 $ & $       0.06 $ & $       0.46 $ \\
CXOMP J005630.6-273141 & $      -0.71 $ & $      -0.46 $ & $      -0.22 $ & $      -0.54 $ & $      -0.30 $ & $      -0.10 $ & $       0.16 $ & $       0.42 $ & $       0.76 $ \\
CXOMP J005633.8-272954 & $      -0.73 $ & $      -0.59 $ & $      -0.47 $ & $      -0.60 $ & $      -0.44 $ & $      -0.30 $ & $       0.33 $ & $       0.50 $ & $       0.70 $ \\
CXOMP J005644.3-274936 & $      -0.60 $ & $      -0.29 $ & $       0.04 $ & $      -1.08 $ & $      -0.64 $ & $      -0.28 $ & $      -0.08 $ & $       0.22 $ & $       0.54 $ \\
CXOMP J005645.6-272444 & $      -0.68 $ & $      -0.35 $ & $      -0.10 $ & $      -0.76 $ & $      -0.38 $ & $      -0.06 $ & $      -0.10 $ & $       0.20 $ & $       0.52 $ \\
CXOMP J005645.9-275522 & $      -0.66 $ & $      -0.53 $ & $      -0.40 $ & $      -0.46 $ & $      -0.33 $ & $      -0.22 $ & $       0.34 $ & $       0.51 $ & $       0.72 $ \\
\enddata
\tablecomments{
Col. (1): ChaMP source name given by right ascension
in units of hours, minutes, and seconds,
and declination in units of degrees, arcminutes, and arcseconds in J2000.
Col. (2): the $68\%$ lower limit of hardness ratio HR.
Col. (3): the mean of hardness ratio (HR$=$(Hc$-$Sc)$/$(Hc$+$Sc)).
Col. (4): the $68\%$ upper limit of hardness ratio HR.
Col. (5): the $68\%$ lower limit of color C21.
Col. (6): the mode of X-ray color C21 ($=log$(S$_{1}/$S$_{2}$)).
Col. (7): the $68\%$ upper limit of color C21.
Col. (8): the $68\%$ lower limit of color C32.
Col. (9): the mode of X-ray color C32 ($=log$(S$_{2}/$H)).
Col. (10): the $68\%$ upper limit of color C32.
For detail information about photometry and color determinations, see \S 3.2.
See Table \ref{tbl-ebands} for definition of energy bands.
The complete version of this table is in the electronic edition of
the $Astrophysical~Journal~Supplement$. The printed edition contains only a sample.
}
\end{deluxetable}
\end{center}
\clearpage

\clearpage
\begin{center}
\begin{deluxetable}{lc|crrrr|crrrr}
\tablecaption{Statistical Summary of the ChaMP X-ray Point Sources \label{tbl-summary}}
\tablewidth{0pt}
\tabletypesize{\footnotesize}
\tablecolumns{12}\tablehead{
\colhead{}&
\colhead{}&
\multicolumn{5}{c}{ALL} &
\multicolumn{5}{c}{$S/N>2.0$} \\ 
\colhead{}&
\colhead{Band}&
\colhead{Number}&
\colhead{Min}&
\colhead{Max}&
\colhead{Median}&
\colhead{Mean}&
\colhead{Number}&
\colhead{Min}&
\colhead{Max}&
\colhead{Median}&
\colhead{Mean}\\
\colhead{}&
\colhead{(1)}&
\colhead{(2)}&
\colhead{(3)}&
\colhead{(4)}&
\colhead{(5)}&
\colhead{(6)}&
\colhead{(7)}&
\colhead{(8)}&
\colhead{(9)}&
\colhead{(10)}&
\colhead{(11)}
}
\startdata
Count\tablenotemark{a} &     B &       6755 &         0.04 &     40535.59 &        21.03 &        68.05 &        5407 &         8.00 &     40535.59 &        28.14 &        83.49 \\ 
 &     S &       6696 &         0.00 &     38117.52 &        15.46 &        54.70 &        4561 &         7.94 &     38117.52 &        25.66 &        77.89 \\ 
 &     H &       6221 &         0.00 &     11604.93 &         5.87 &        15.12 &        1914 &         8.28 &     11604.93 &        18.02 &        39.16 \\ 
 &    Bc &       6736 &         0.20 &     39760.98 &        20.53 &        65.33 &        4950 &         9.53 &     39760.98 &        30.05 &        86.45 \\ 
 &    Sc &       6634 &         0.01 &     36010.96 &        13.52 &        48.23 &        4245 &         7.94 &     36010.96 &        24.27 &        72.73 \\ 
 &    Hc &       6401 &         0.00 &     13624.92 &         7.16 &        18.85 &        2406 &         8.33 &     13624.92 &        18.88 &        42.24 \\ 
\hline
Flux\tablenotemark{b} &     B &       6755 &         0.01 &      6484.14 &         7.97 &        23.10 &        5407 &         0.80 &      6484.14 &        10.04 &        27.52 \\ 
 &     S &       6696 &         0.00 &      3482.36 &         3.92 &        12.19 &        4561 &         0.45 &      3482.36 &         5.96 &        16.71 \\
\enddata
\tablecomments{
Col. (1): X-ray energy band.
Col. (2)-(6): the statistical properties of all ChaMP X-ray point sources.
Col. (7)-(11): the statistical properties of sources with $S/N>2.0$.
}
\tablenotetext{a}{Source Counts}
\tablenotetext{b}{Source Flux with $\Gamma_{ph}=1.7$ in units of $1\times10^{-15}~erg~cm^{-2}~sec^{-1}$}
\tablenotetext{c}{Effective Exposure Time in units of $ksec$}
\tablenotetext{d}{Off axis angle in units of arcminutes}
\end{deluxetable}
\end{center}
\clearpage

\clearpage

\plotone{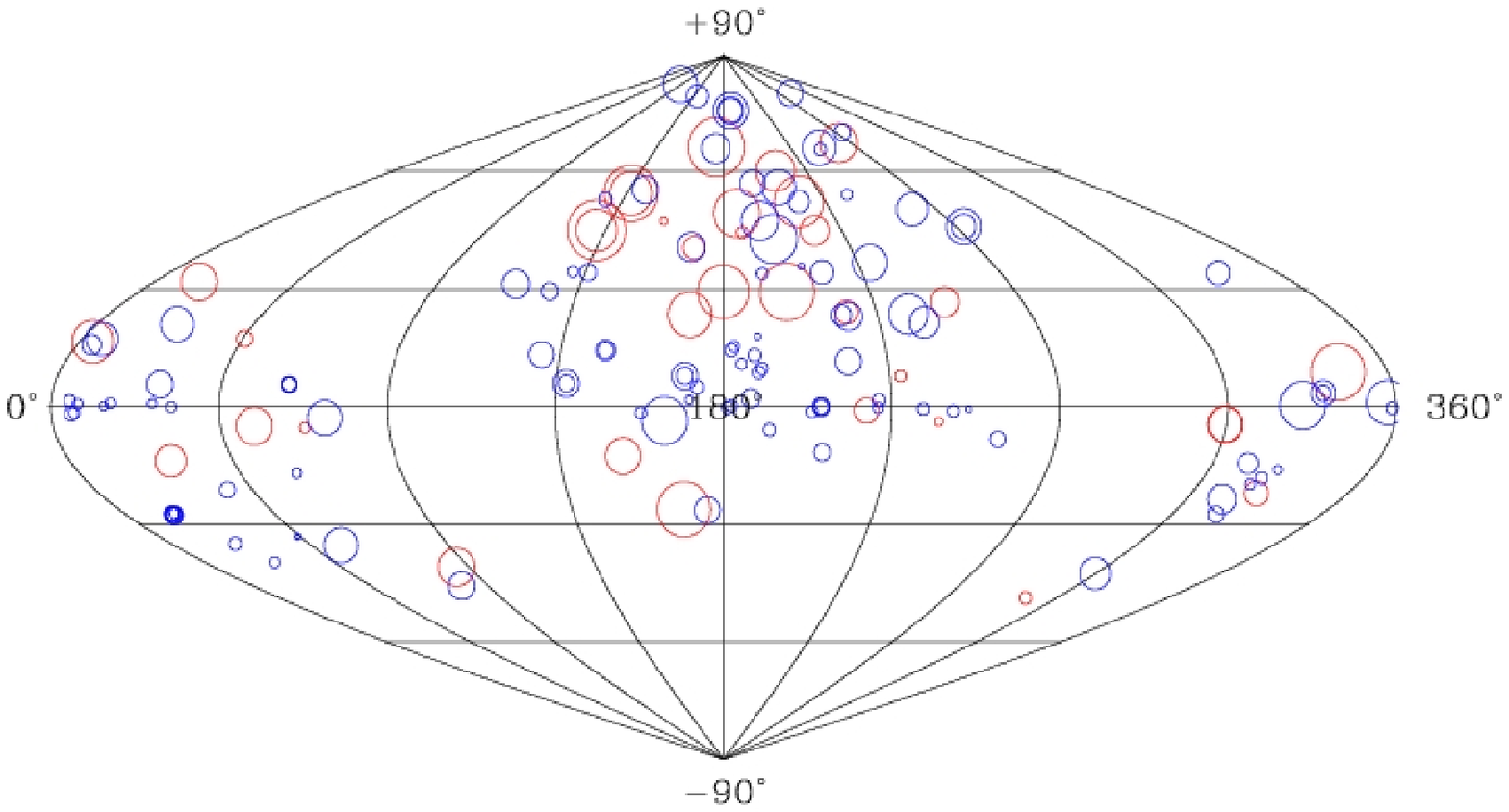}
\figcaption[f1.eps]{
The location of 149 ChaMP fields in equatorial
coordinates.
Red circles represent ACIS-I at the aim point, and blue circles ACIS-S.
The circle size crudely indicates the Chandra exposure time,
ranging from 0.9 to 124 ksecs.
The ChaMP fields are uniformly distributed over the celestial space except 
(by selection) the Galactic plane region.
\label{fig-pos}}
\clearpage

\plotone{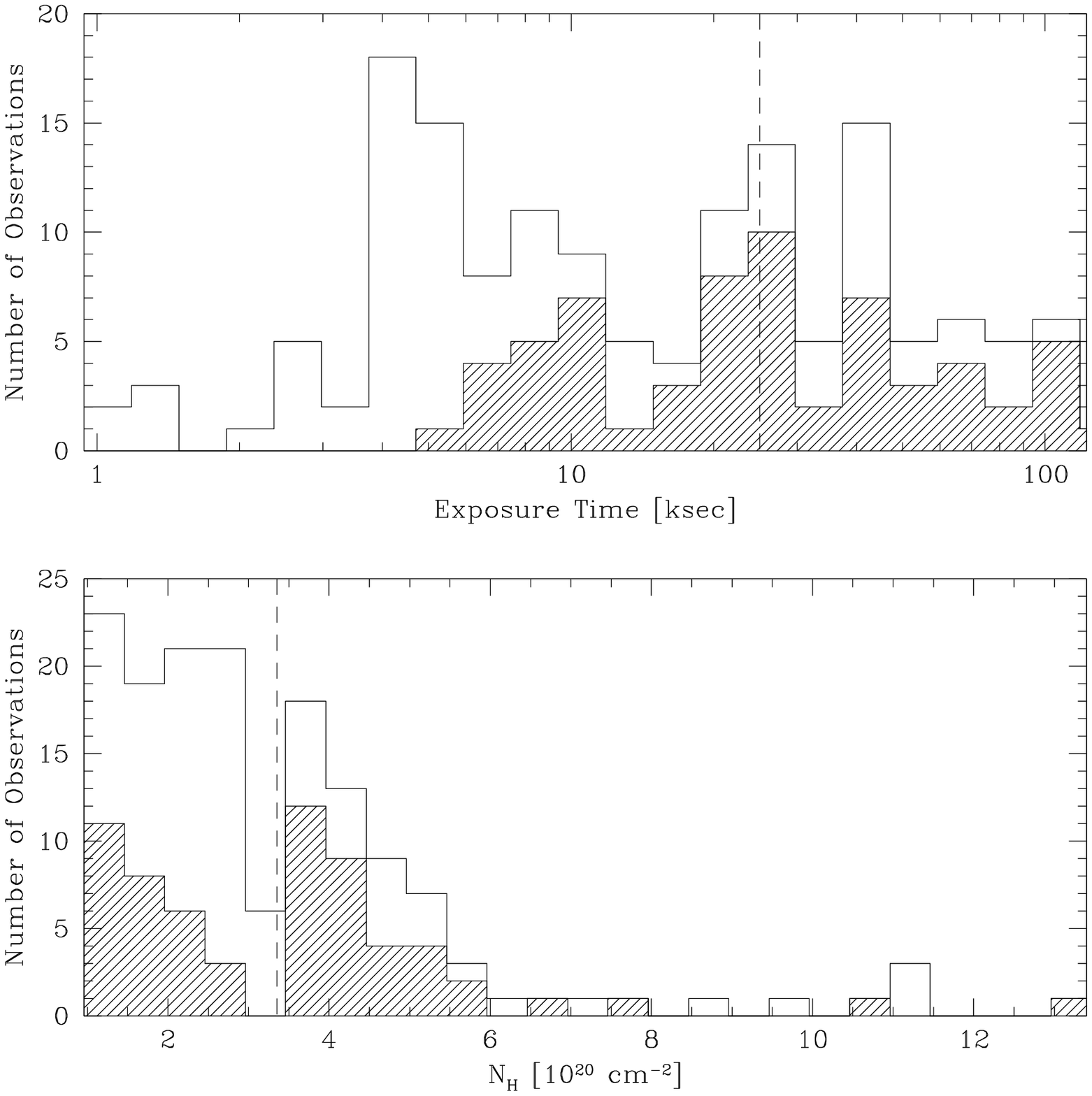}
\figcaption[f2.eps]{
The number distribution of effective exposure times ($top$)
and Galactic extinction, $N_{H}$ ($bottom$) of the 149 ChaMP observations.
The dashed line indicates the mean of each number distribution.
The mean exposure time of the ChaMP is $\sim25$ $ksec$
and the mean Galactic extinction in the ChaMP is
$N_{H}\sim(3.4\pm2.2) \times 10^{20}~cm^{-2}$.
The ChaMP fields range from short
to long exposure times and their Galactic absorption is lower than 
the Galactic plane ($N_{H}\sim 10^{22}\sim10^{23}~cm^{-2}$).
The number distributions of the 62 ChaMP fields from Paper I
are displayed as shaded histograms.
\label{fig-expnh}}
\clearpage

\plotone{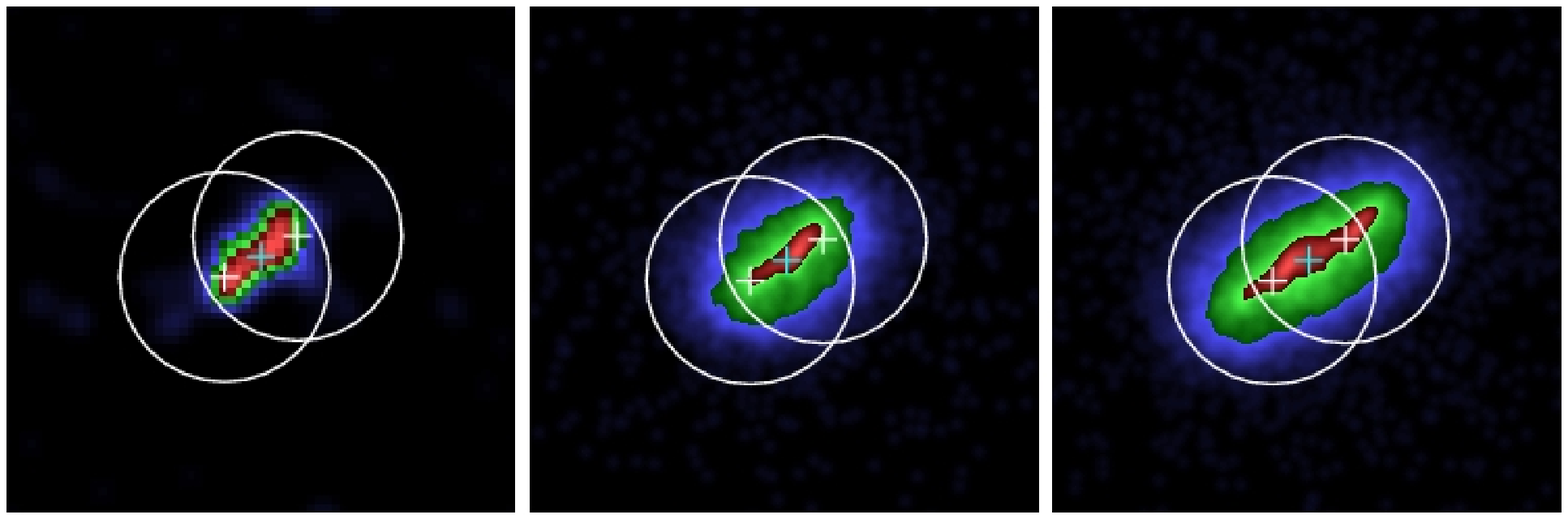}
\figcaption[f3.eps]{
The large overlapping double X-ray sources detected by $wavdetect$ ($left$),
the PSF image at their median location ($middle$),
and the combined image of two PSFs at each X-ray source position ($right$).
The white crosses and circles mark the positions and sizes of the double sources.
The cyan crosses indicate the median location of the double sources.
Circle size corresponds to the source extraction region used in $xapphot$.
The shape of the detected overlapping X-ray sources ($left$) are similar to
the shape of the single PSF ($middle$), rather than double PSFs ($right$).
We conclude that the overlapping X-ray sources are spurious doubles
due to the elongated PSF shape and sub-structures in the PSF.
The PSF images are generated by the $ChaRT$ tool in the CIAO package 
and MARX package.
The size of all images is $\sim 30\arcsec \times 30\arcsec$ and
all images are aligned with the world coordinate system.
\label{fig-psfeffect}}
\clearpage

\plotone{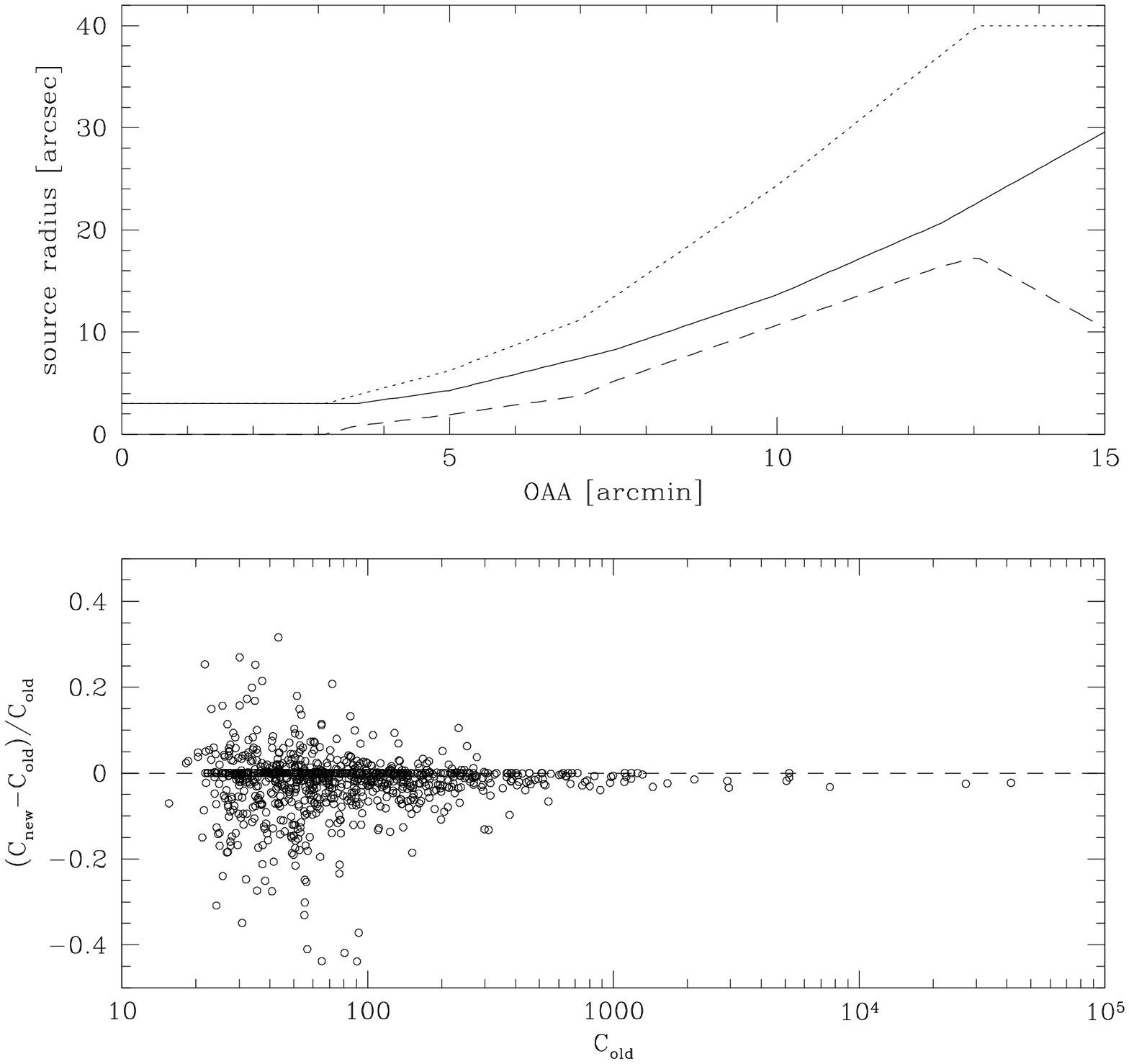}
\vspace{-10mm}
\figcaption[f4.eps]{
$top.$ Source extraction radius as a function of off-axis angle.
The dotted line represents the source radius from the old version of the CIAO CALDB
used in the previous ChaMP catalog (Paper I). The solid
line represents the source radius from the latest version of the CIAO CALDB
used in this catalog. The source size is limited to a minimum radius of
$3\arcsec$ and a maximum of $40\arcsec$.
The difference between the two radii is plotted as a dashed line.
$bottom.$
The difference between the source counts in this and the previous catalog
as a function of the previous counts.
On average the reduced source radius yields source counts 
lower by $\sim2\pm7\%$ in this catalog.
The dashed line indicates the zero difference level.
\label{fig-prop}}
\clearpage

\plotone{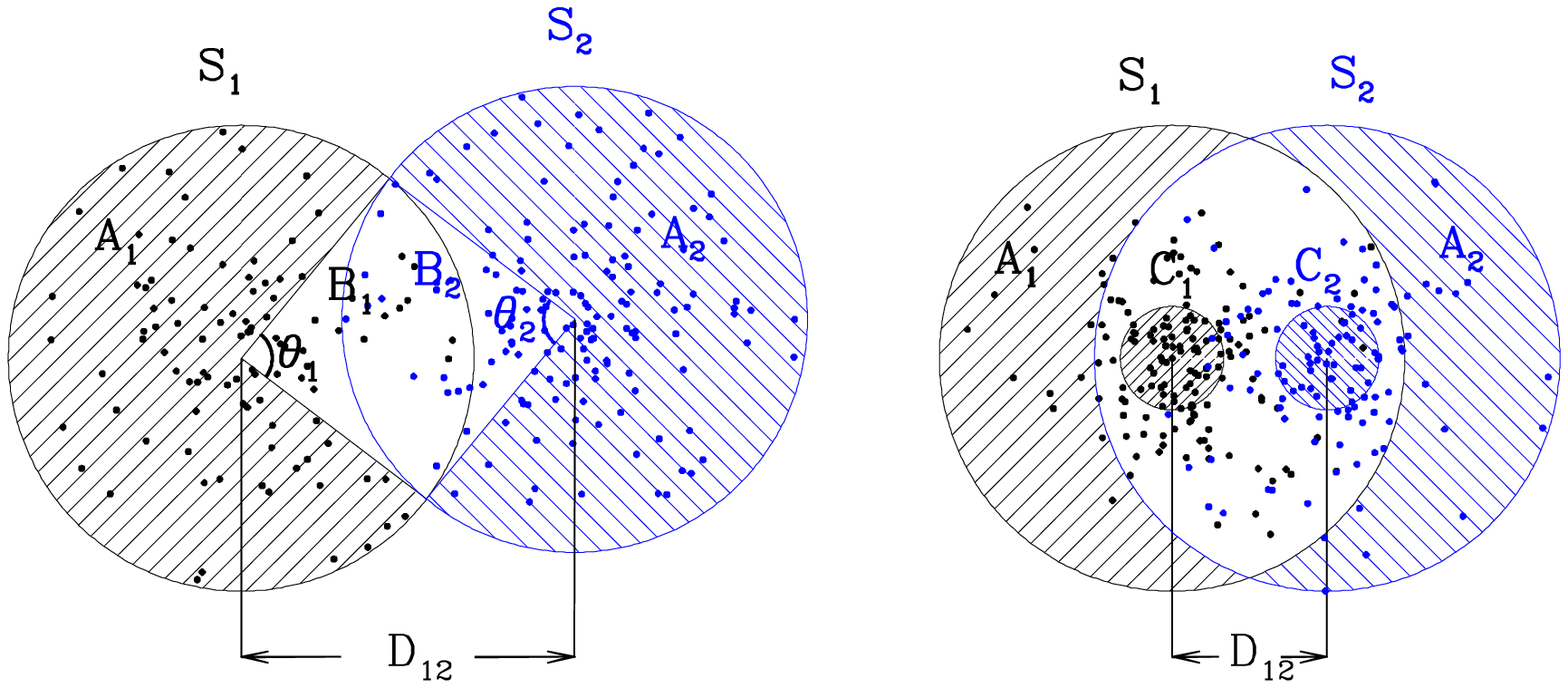}
\vspace{-20mm}
\figcaption[f5.eps]{
Schematic diagrams of the two types of overlapping X-ray source pairs,
small ($left$) and large ($right$).
$S_1$ and $S_2$ represent the source1 and source2, respectively.
$A_1$ and $A_2$ represent the independent area of each source.
$\theta_1$ and $\theta_2$ represent the angles covered by sectors
$B_1$, and $B_2$ of overlapping regions of two sources, respectively.
$C_1$ and $C_2$ represent the core regions of large overlapping sources,
respectively. $D_{12}$ represents the distance between $S_{1}$ and $S_{2}$.
The points are simulated X-ray events assuming a $\beta$ model for the
event distribution.
For the small overlapping case, only the photons in regions $A_1$ and $A_2$ are
used to determine the correction, while the photons in regions $A's$ and $C's$
are used in the large overlapping case. The detailed correction method 
of $xapphot$ is described in the text.
\label{fig-small_large}}
\clearpage

\plotone{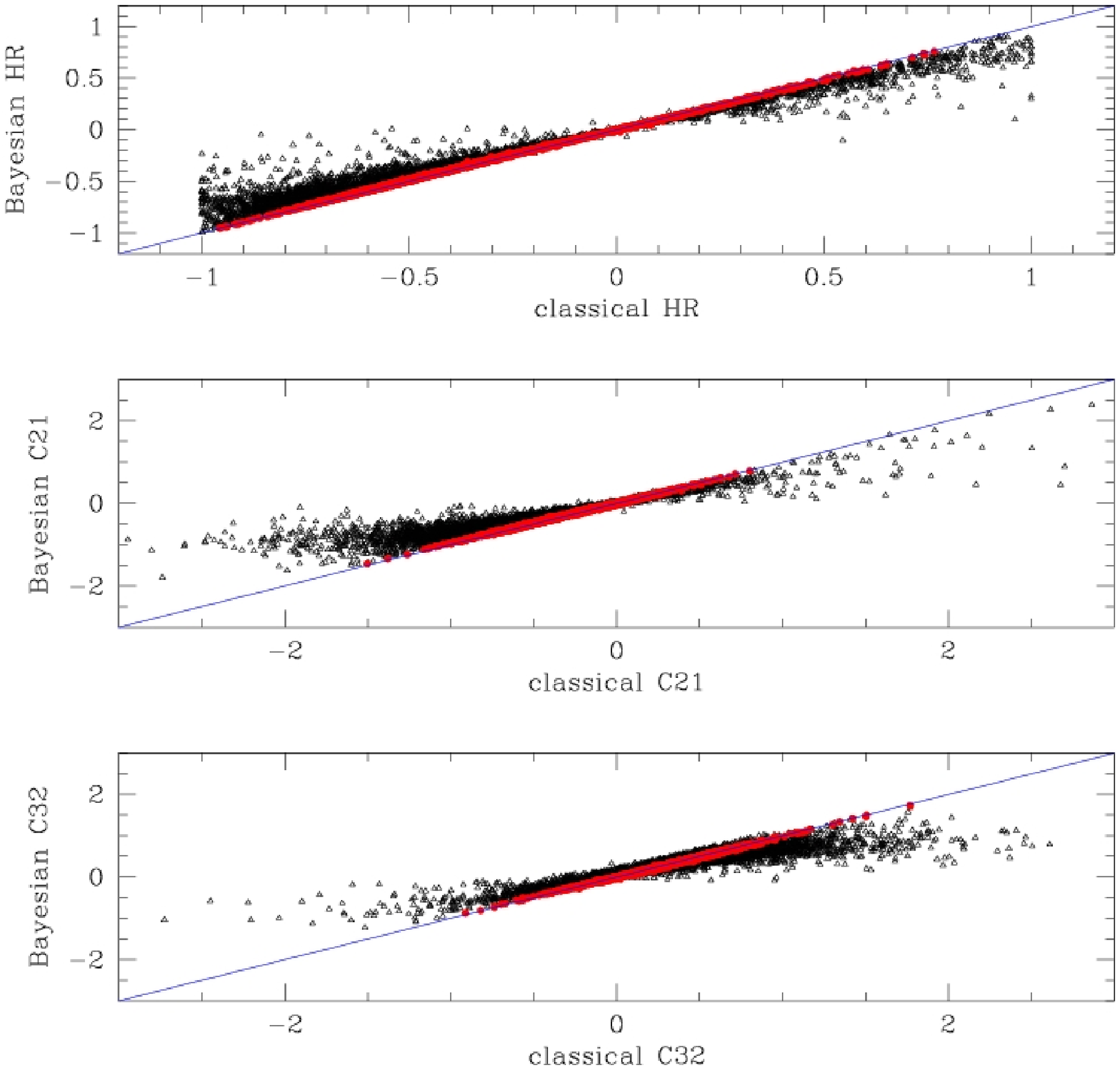}
\vspace{-10mm}
\figcaption[f6.eps]{
Comparison of a classical with a Bayesian method for
HR ($top$), C21 ($middle$), and C32 ($bottom$). 
The sources with $net~counts\geq0$ ($net~counts>0$) for HR (X-ray colors)
in both energy bands are plotted.
The black open triangles represent the sources with $S/N<2$ in 
at least one energy band.
The red closed circles represent the sources with $S/N>2$ in two energy bands.
The blue line represents the line of equality
for the two methods and is shown for illustrative comparison.
For bright sources ($S/N>2$), the HR and X-ray colors from both methods 
agree well; however, for faint sources ($S/N<2$), they  
do not agree, because the classical method, using Gaussian statistics, 
fails to describe the nature of faint sources.
\label{fig-hcolor_final}}
\clearpage

\clearpage

\plotone{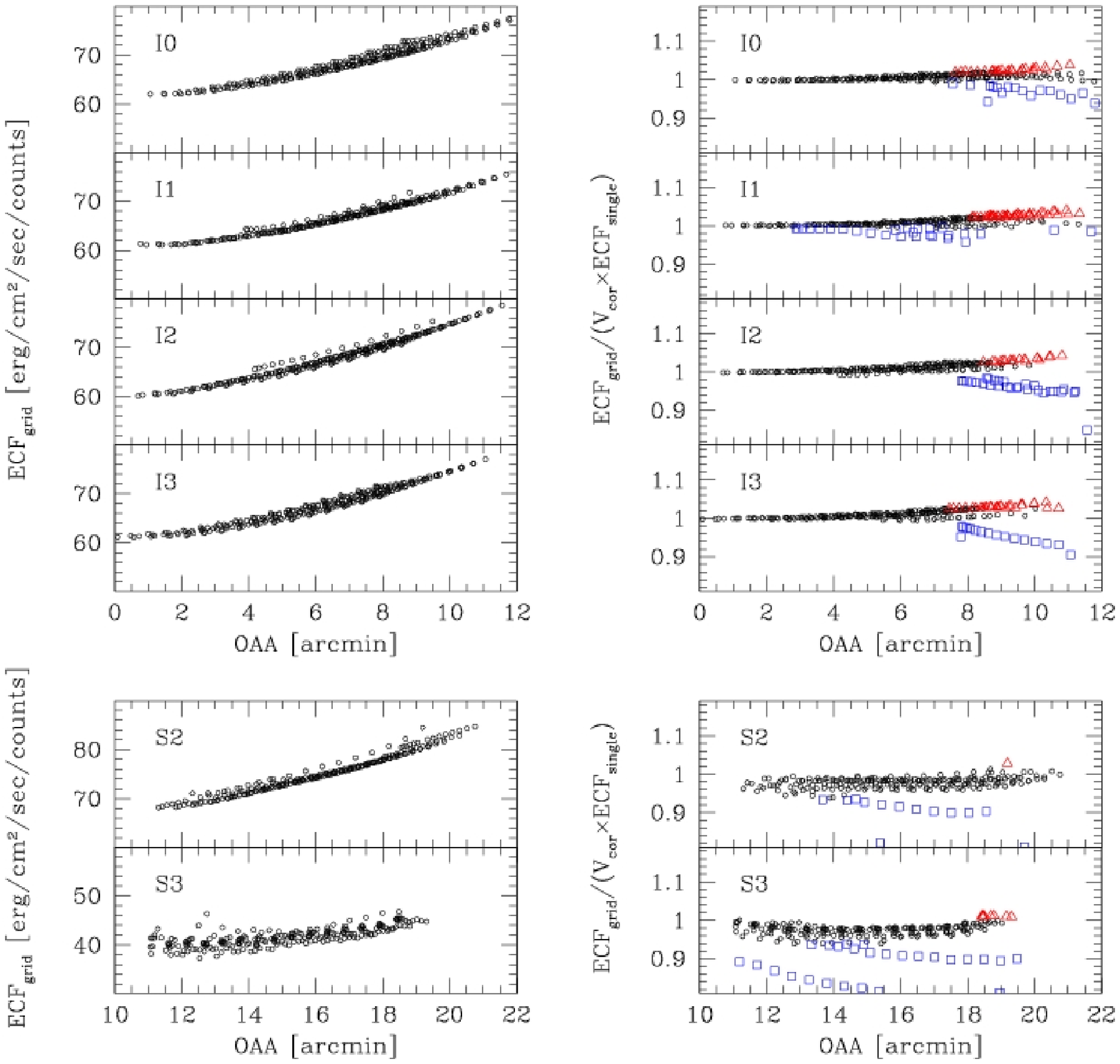}
\vspace{-20mm}
\figcaption[f8.eps]{
$left.$ The spatial variations of the $ECF_{grid}$ in the S band
as a function of off axis angle in each CCD chip
observed with ACIS-I including S2 and S3 CCD chips.
The aim point of this observation is
located on the I3 CCD chip, so the off axis angle of S2 and S3 chips
is large. $ECF_{grid}$ varies by at most $\sim 25\%$
depending on position in each CCD chip.
$right.$ The ratios of $ECF_{grid}$
to the vignetting corrected $ECF_{single}$. 
Blue squares are caused by the exposure map defects such as a CCD
chip edge or bad pixel strip effect. Red triangles are caused by
lower quantum efficiency at larger off axis angles.
The mean ratio is $1.00\pm0.02$ in the I0-I3 chips
and $0.98\pm0.02$ and $0.97\pm0.03$ in the S2 and S3 chips.
\label{fig-ecf_cor_vari}}
\clearpage

\plotone{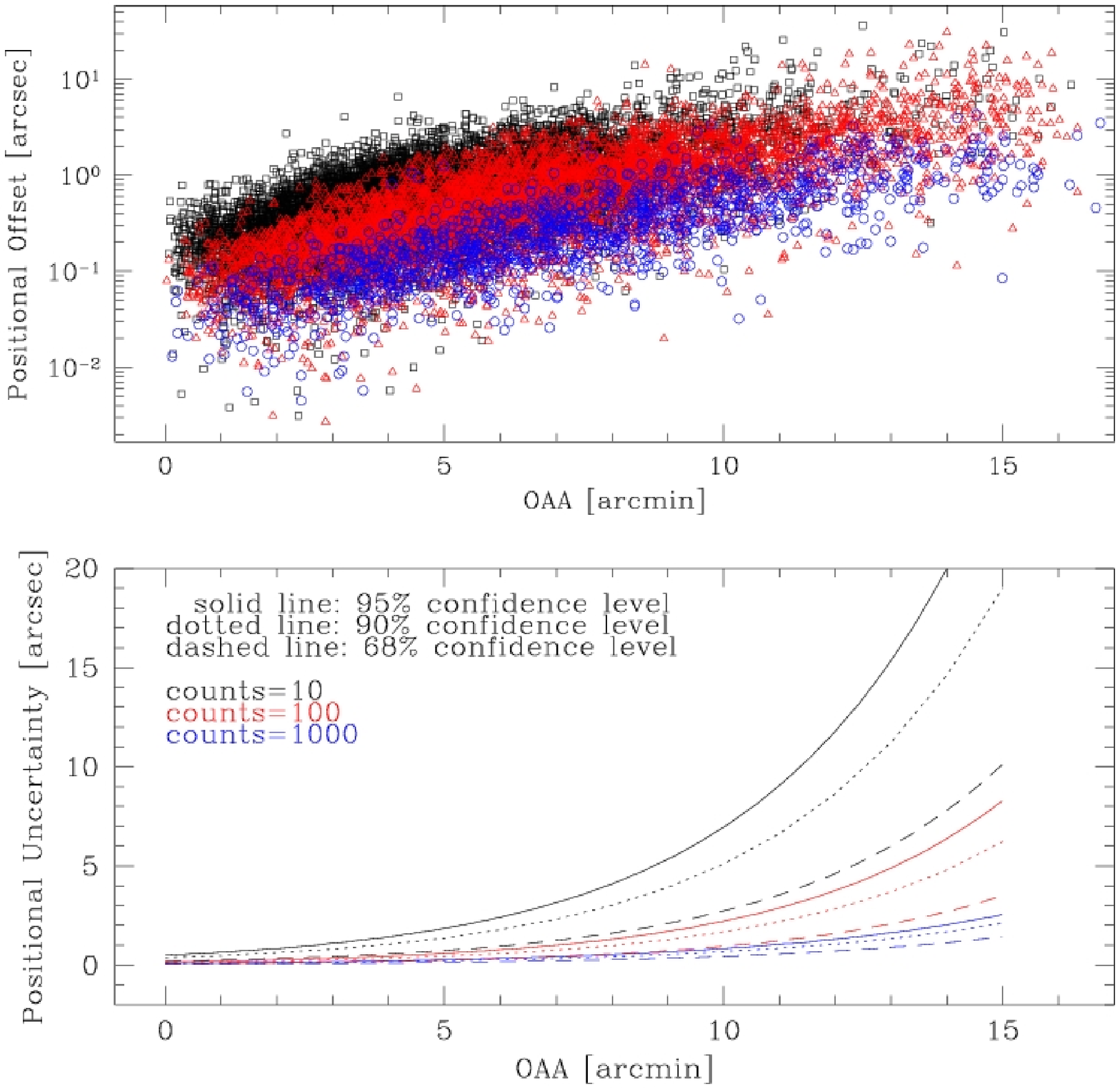}
\vspace{-15mm}
\figcaption[f9.eps]{
$top.$ The positional offset
of the artificial sources in three source count ranges
as a function of off-axis angle.
The black squares, red triangles, and blue circles represent
artificial sources with count ranges of $0<logC<1.2$,
$1.2<logC<2$, and $logC>2$, respectively.
The positional offset exponentially increases
with off axis angle, and decreases as the source counts increase 
with a power law form.
$bottom.$ The positional uncertainties from the derived equations
(see equation (12), (13), and (14) in \S 4.2.1)
as a function of off axis angle for  
10 source counts ($black$), 100 source counts ($red$), 
and $1,000$ source counts ($blue$), respectively.
The solid, dotted, and dashed lines represent the positional
uncertainty at $95\%$, $90\%$, and $68\%$ confidence levels,
respectively. 
\label{fig-poserr_off_eq}}
\clearpage

\plotone{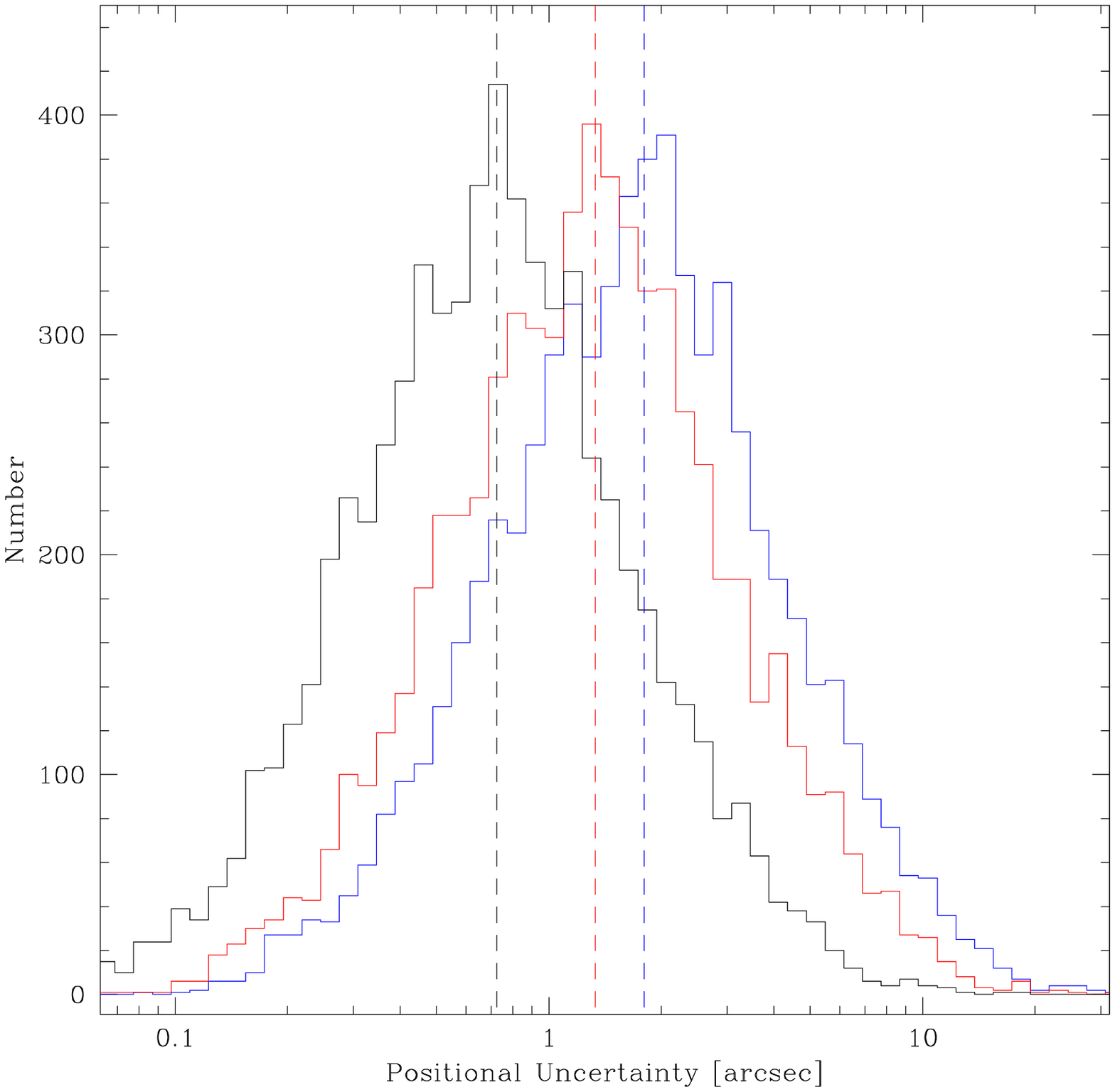}
\figcaption[f10.eps]{
The number distributions of positional uncertainty of
all ChaMP X-ray point sources estimated from our empirical 
positional uncertainty equations.
Black, red, and blue histograms show the $68\%$, $90\%$, and $95\%$
confidence level of positional uncertainty distributions, respectively.
The median positional uncertainties for all sources
are plotted as dashed lines and
shown at $0.7\pm0.45\arcsec$, $1.3\pm0.8\arcsec$, 
and $1.8\pm1.1\arcsec$, respectively.
\label{fig-poserr}}
\clearpage

\plotone{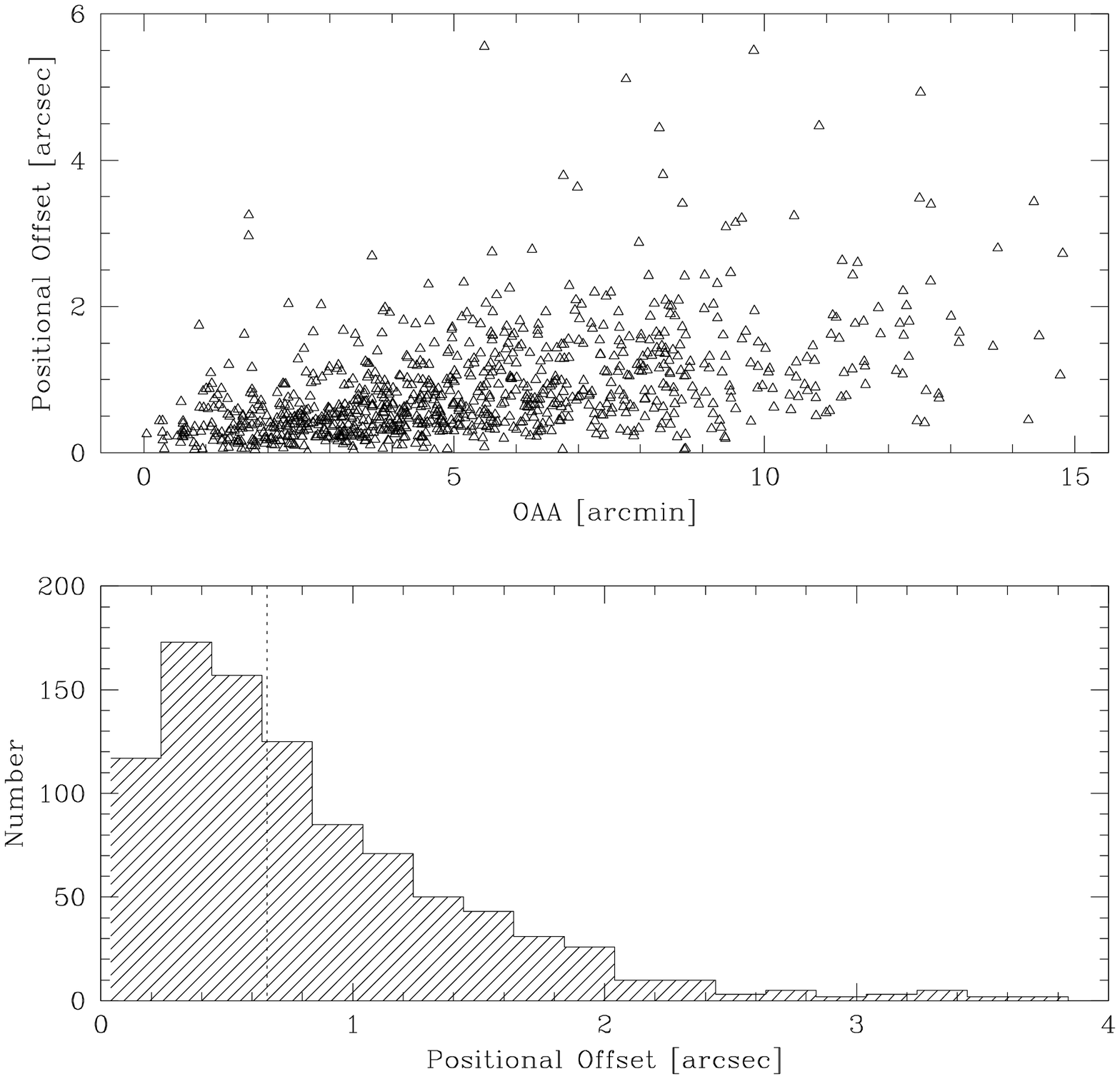}
\figcaption[f11.eps]{
$top.$ Positional offset between the matched ChaMP and $SDSS$ sources
as a function of off-axis angle.
$bottom.$ The number distribution of the positional offset between
the matched ChaMP and the $SDSS$ sources.
The median positional offset is $0.7\pm0.4\arcsec$ and is denoted
by the dotted line.
\label{fig-aspcor}}
\clearpage

\plotone{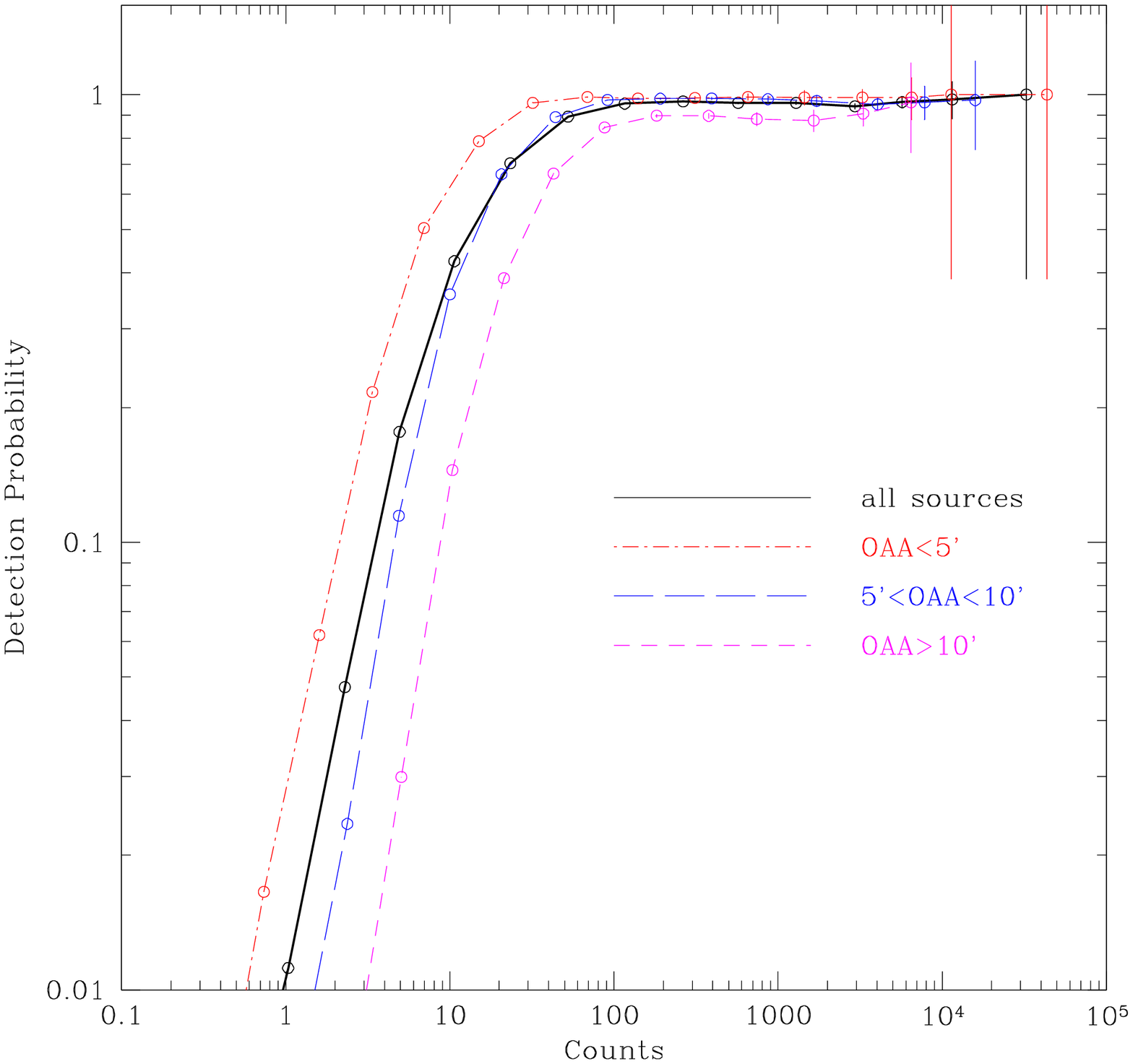}
\figcaption[f12.eps]{
The detection probability of a source in the ChaMP 
catalog as a function of B band counts.
The thick solid line represents the detection probability of all sources.
The red dot dashed, blue long dashed, 
and magenta short dashed lines denote the detection probability
of sources located at off axis angles, 
$OAA<5\arcmin$, $5\arcmin<OAA<10\arcmin$,
and $OAA>10\arcmin$, respectively.
As expected, the detection probability decreases as 
the off axis angle increases.
\label{fig-detectp_c_off}}
\clearpage

\plotone{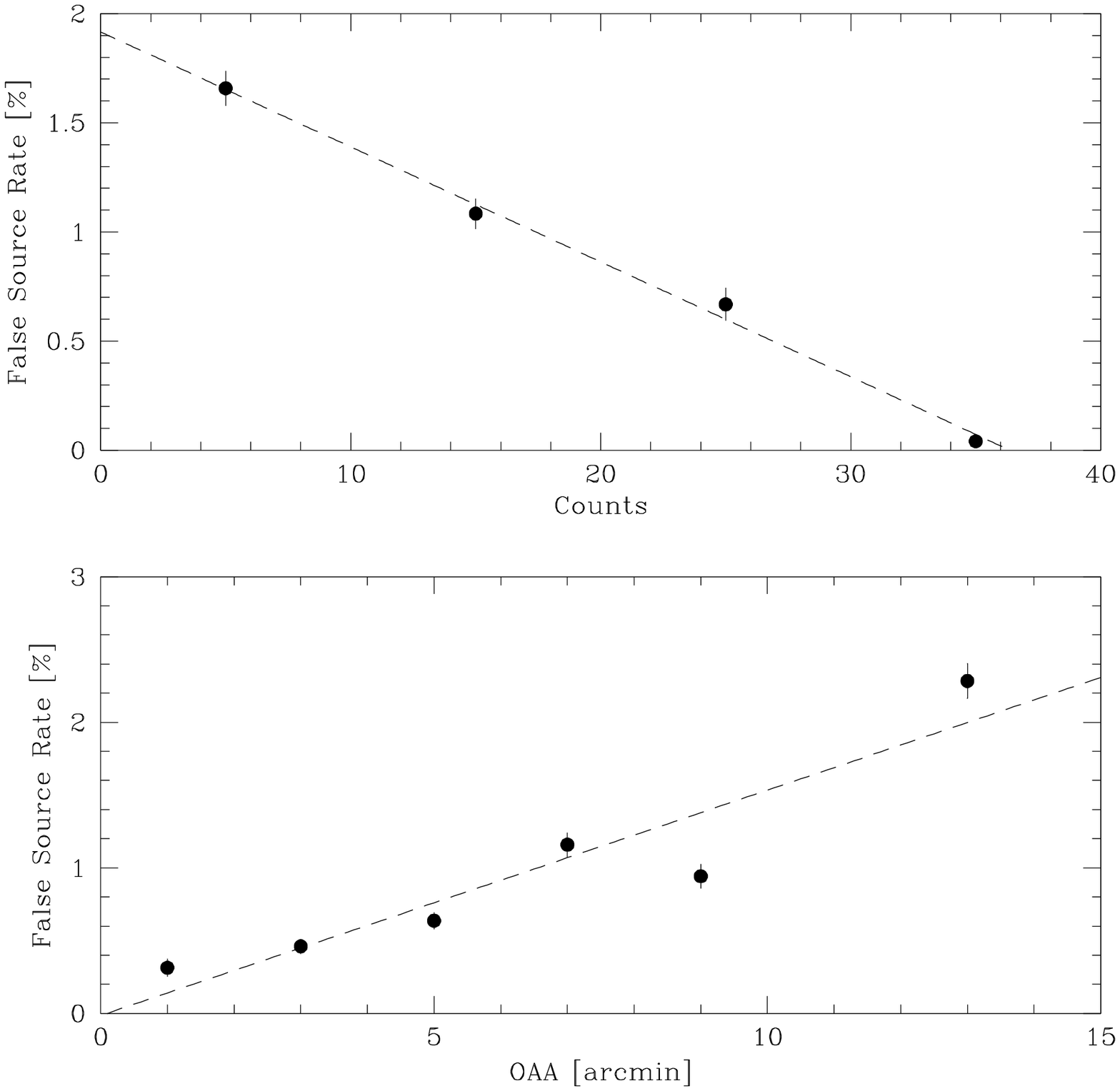}
\figcaption[f13.eps]{
The false source detection rate as a function of source counts in the 
B band extracted by $wavdetect$ ($top$) and off axis angle 
($bottom$).
$\sim 1\%$ of the total detected sources are spurious sources
and $80\%$ of spurious sources have counts less than $\sim 30$.
The false source detection rate increases as the source counts decrease
and as the off axis angle increases.
The dashed lines indicate the best linear least square fit results
(see \S 4.3.2).
\label{fig-false_count}}
\clearpage

\plotone{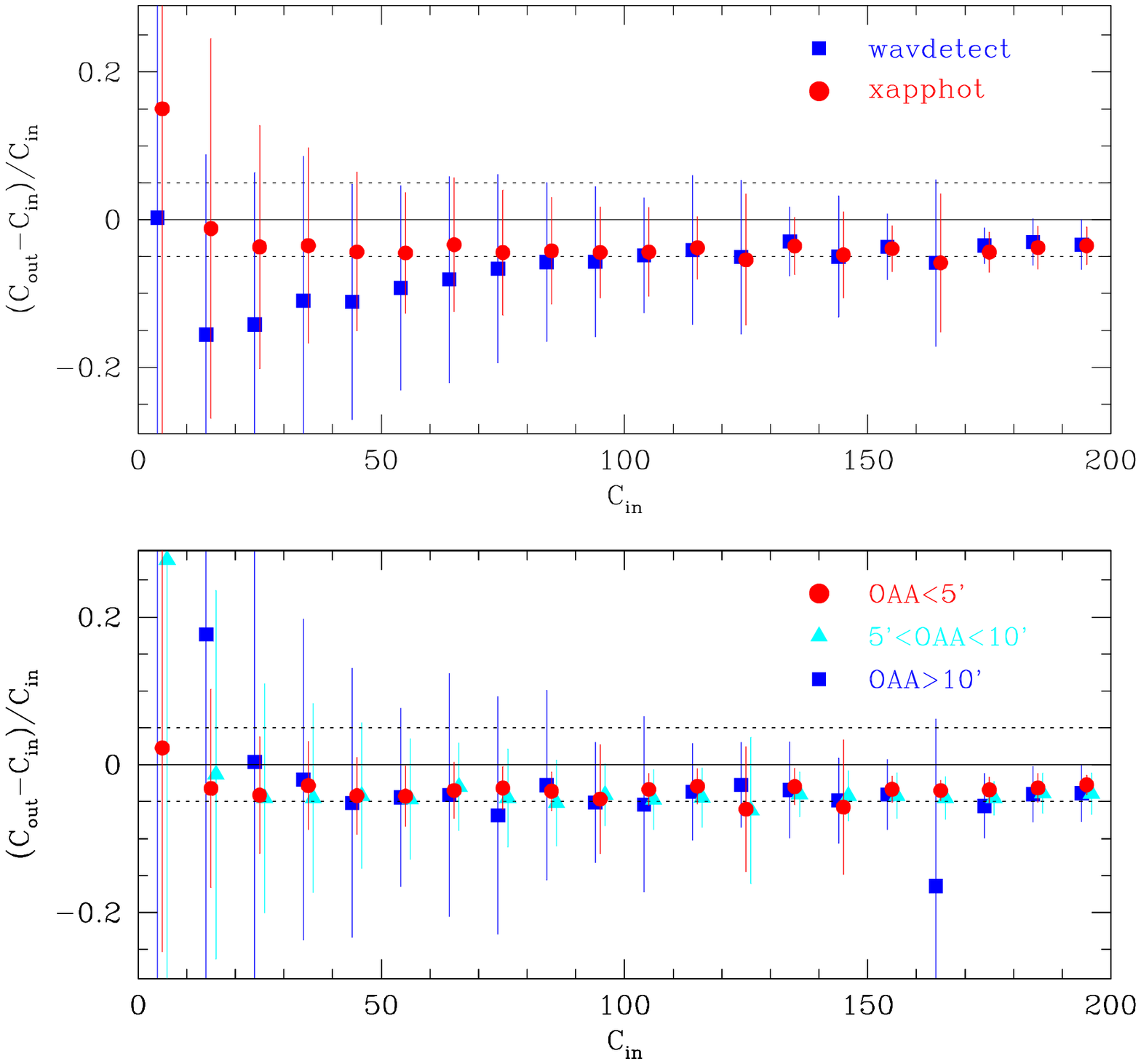}
\vspace{-15mm}
\figcaption[f14.eps]{
The count recovery rate as a function of input counts.
$top.$ Comparison of the count recovery rate 
from $xapphot$ ($red$ $circles$) 
with that from $wavdetect$ ($blue$ $squares$).
Both $xapphot$ and $wavdetect$ recover the true counts well
($96\pm1\%$ level for $xapphot$ and $94 \pm 3\%$ level for $wavdetect$),
however, for source counts fainter than $\sim 50$, $wavdetect$
recovers only $87\pm2\%$ of the true counts. 
Note that Eddington bias is visible in the first points of both 
$wavdetect$ and $xapphot$ count recovery rates.
$bottom.$ 
The count recovery rates of sources with
$OAA<5\arcmin$ ($red$ $circle$), $5\arcmin<OAA<10\arcmin$ ($cyan$ $triangle$), 
and $OAA>10\arcmin$ ($blue$ $square$), respectively.
Source counts are extracted using $xapphot$.
As the off axis angle increases the uncertainty in the 
count recovery rate increases.
\label{fig-xcompmag1}}
\clearpage

\plotone{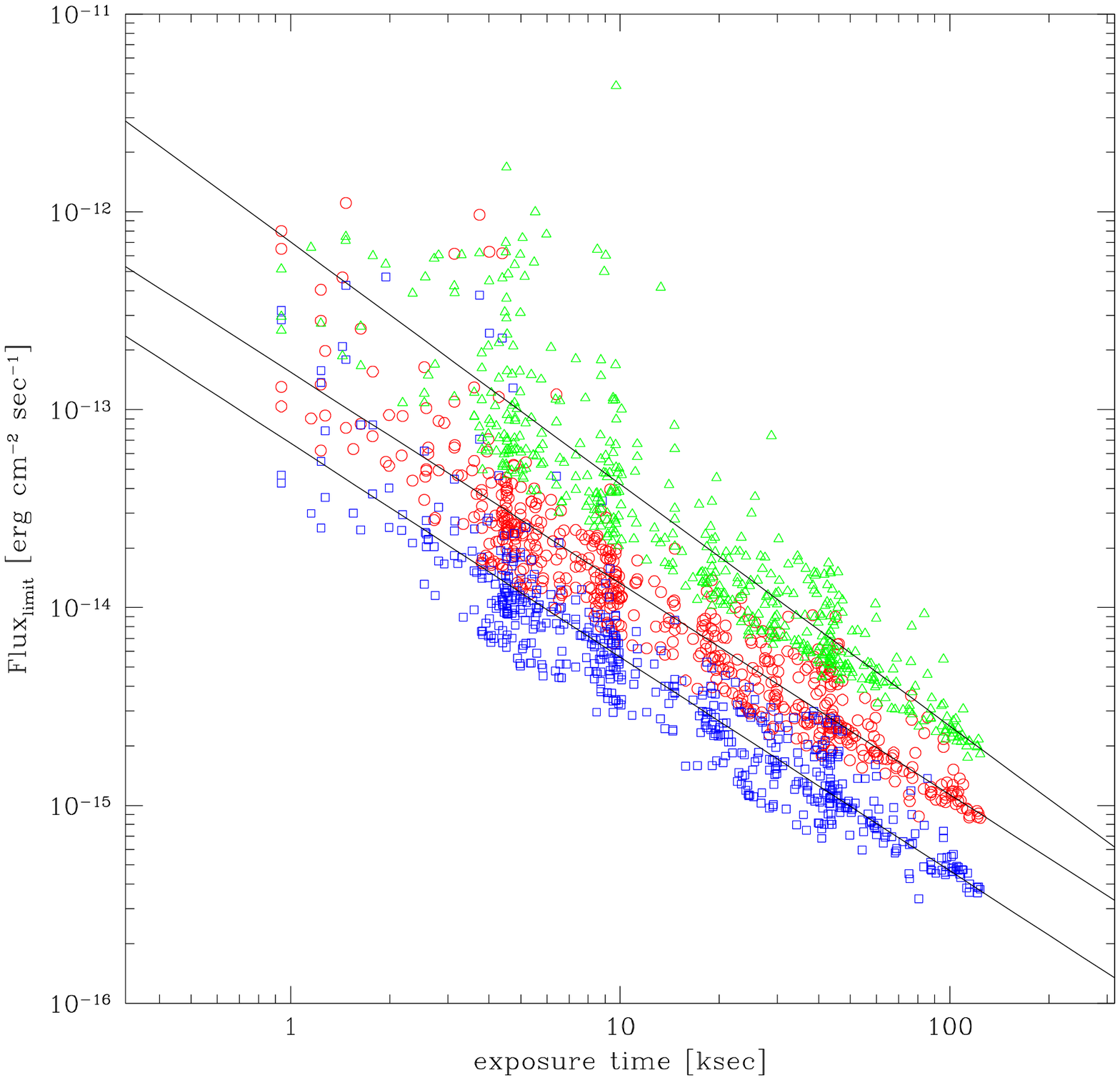}
\vspace{-18mm}
\figcaption[f15.eps]{
The flux limits of detected X-ray point sources in the ChaMP fields
as a function of net exposure time.
Four CCD chips (I0, I1, I2, and I3 for ACIS-I
observations, and I2, I3, S2, and S3 for ACIS-S observations) per ChaMP field
and the 130 ChaMP fields were used. 
Detected artificial X-ray point sources with
$S/N>2.0$ are selected in each CCD chip
and their minimum
flux is defined as the flux limit of that CCD chip.
Red circles, blue squares, and green triangles 
represent the Bc, Sc, and Hc bands, respectively.
The solid lines represent
the best linear least squares fit results in each energy band.
The scatter is caused by the varying sensitivity and detection probability
of each CCD chip and observation. 
A photon index of $\Gamma_{ph}=1.7$ was assumed. 
\label{fig-fluxlim_cband}}
\clearpage

\plotone{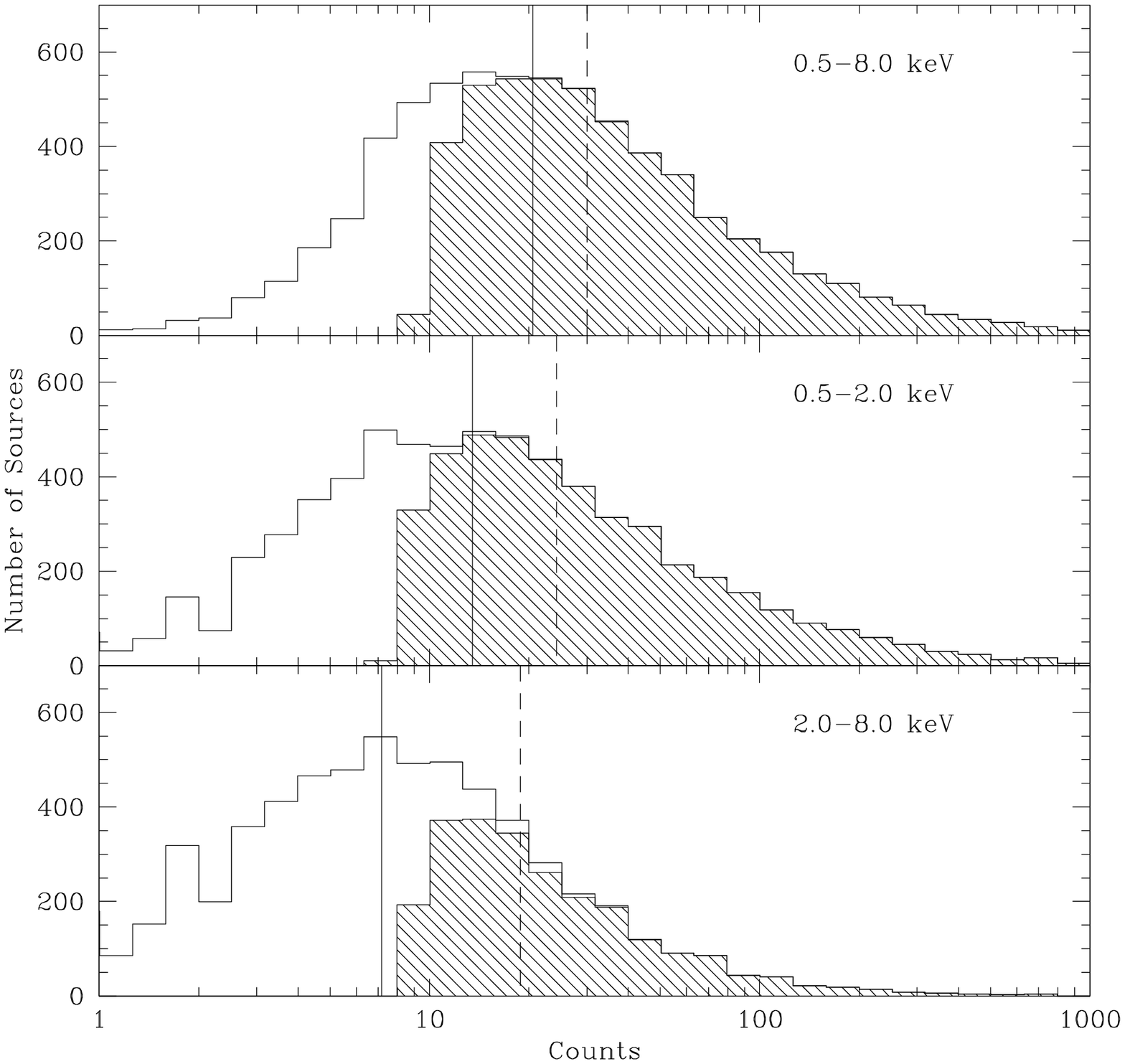}
\figcaption[f16.eps]{
The distributions of source net counts 
in the Bc ($top$), Sc ($middle$), and Hc ($bottom$) bands, respectively.
The open and shade histograms are for all sources and 
sources with $S/N>2.0$, respectively, in each energy band.
The solid and dashed vertical lines indicate the medians of the total
sample and high $S/N$ sample, respectively.
\label{fig-count}}
\clearpage

\plotone{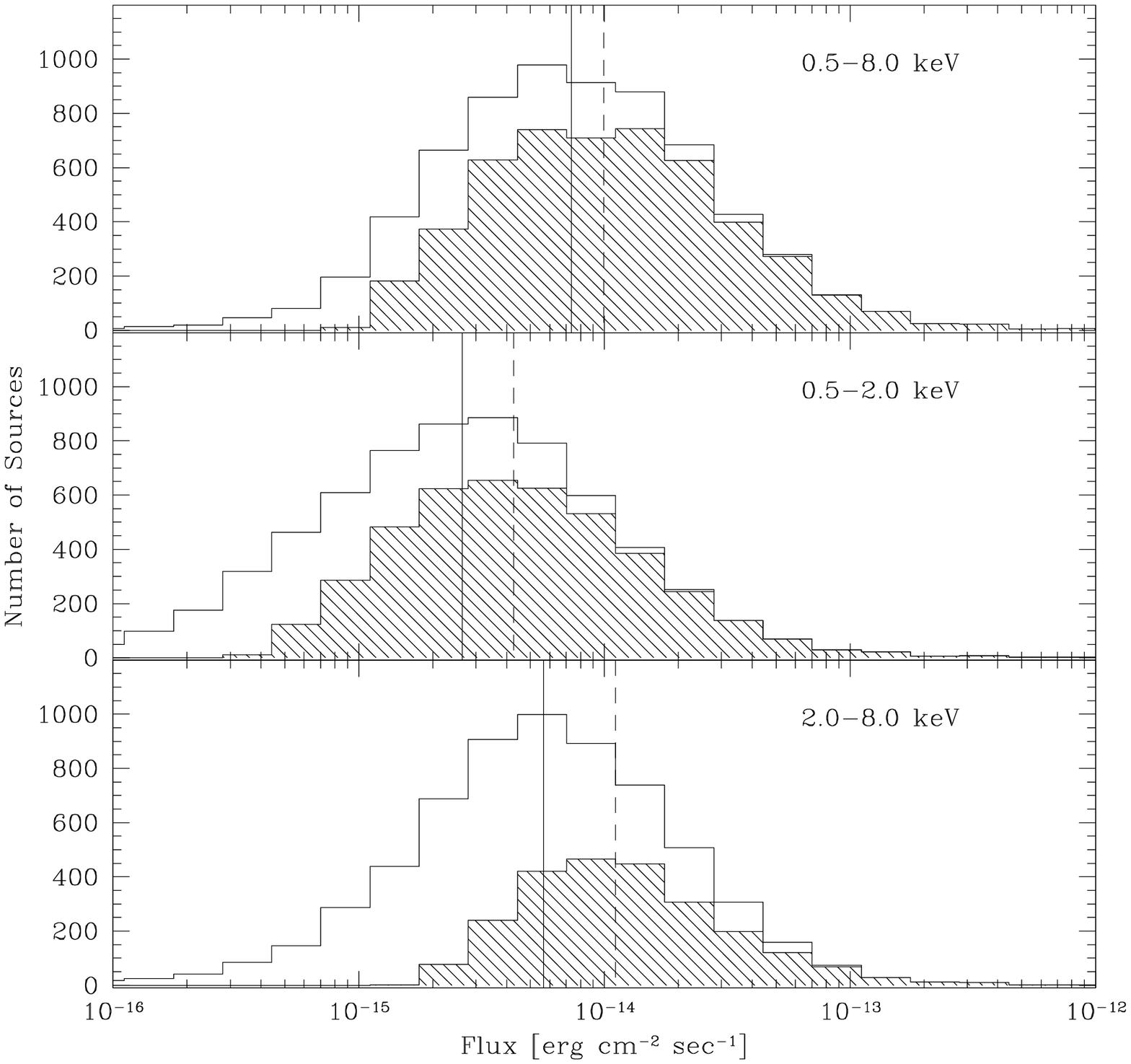}
\vspace{-10mm}
\figcaption[f17.eps]{
Same as Figure \ref{fig-count}, but for flux.
The flux was determined assuming a photon index of $\Gamma_{ph}=1.7$
and Galactic absorption $N_{H}$.
\label{fig-flux}}
\clearpage

\plotone{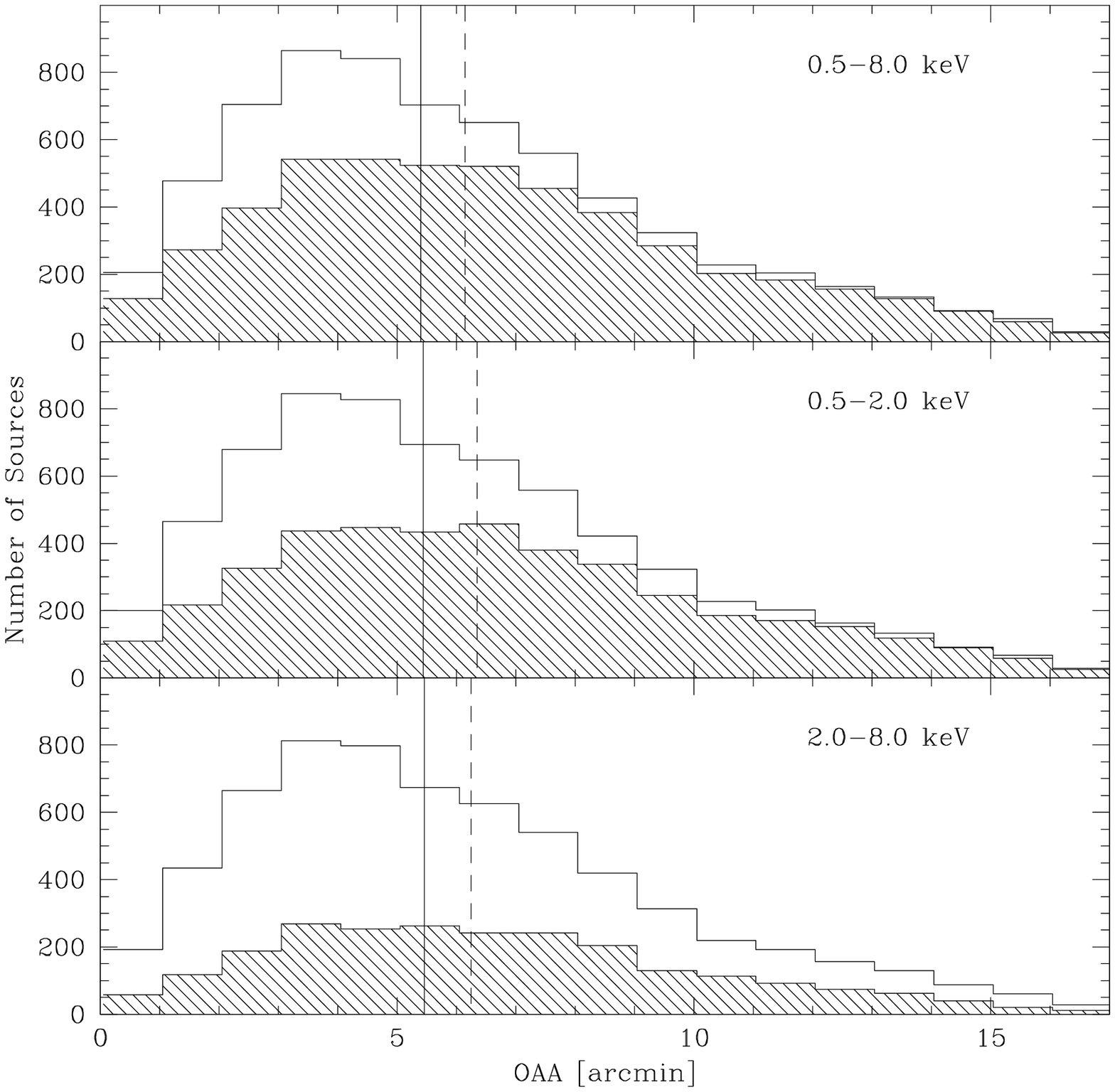}
\figcaption[f18.eps]{
Same as Figure \ref{fig-count}, but for off axis angle.
\label{fig-off}}
\clearpage

\plotone{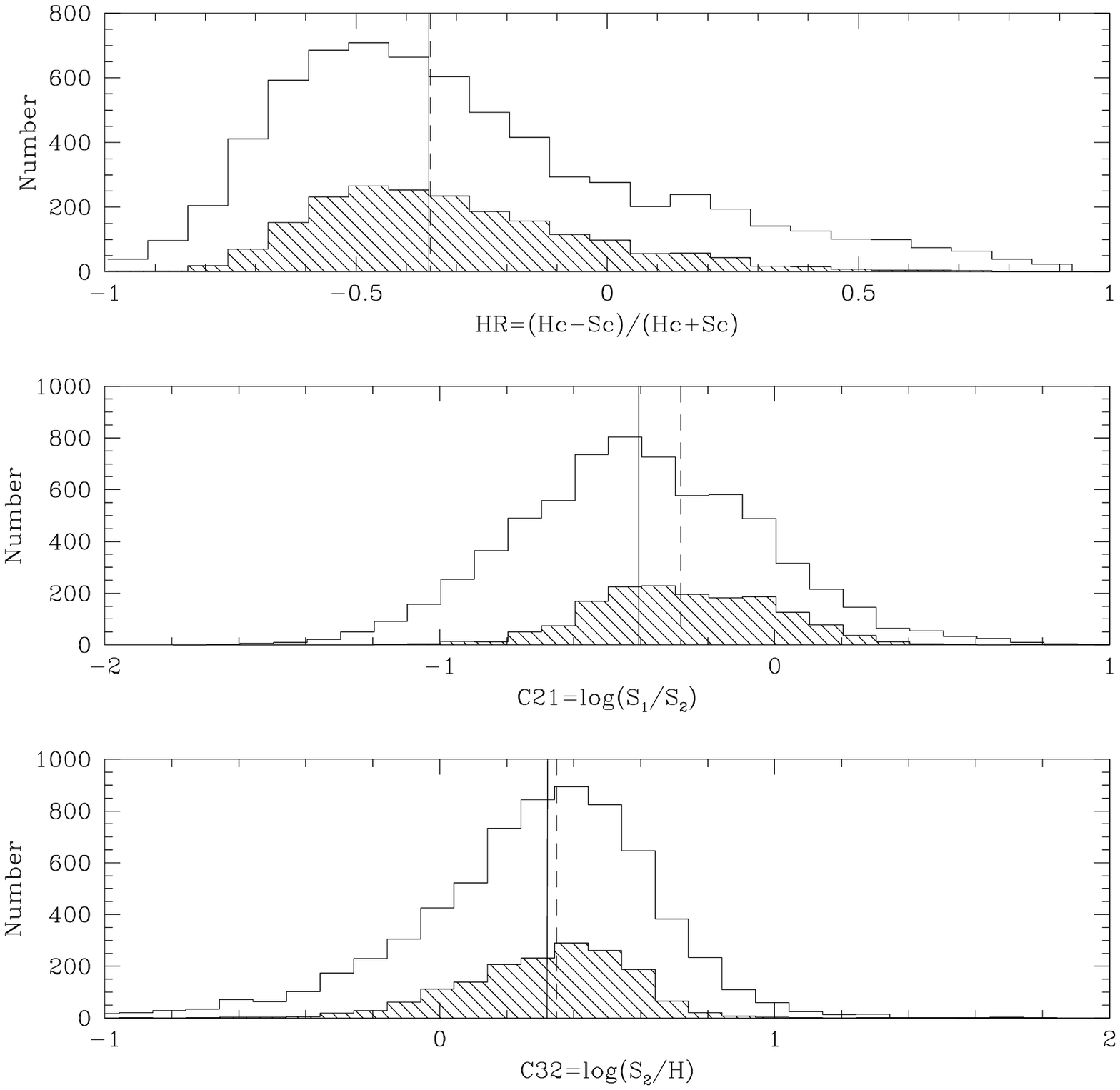}
\vspace{-10mm}
\figcaption[f19.eps]{
The number distribution of mean hardness ratio ($top$), 
the most probable values of the color C21 ($middle$)
and C32 ($bottom$) of ChaMP X-ray point sources.
The open and shaded histograms are for
sources with $S/N\geq0$ and $S/N>2$ in both energy bands, respectively.
This constraint yields a small number of shaded histograms.
The solid and dashed vertical lines indicate the medians of the total
sample and high $S/N$ sample, respectively.
\label{fig-hcolor}}
\clearpage

\plotone{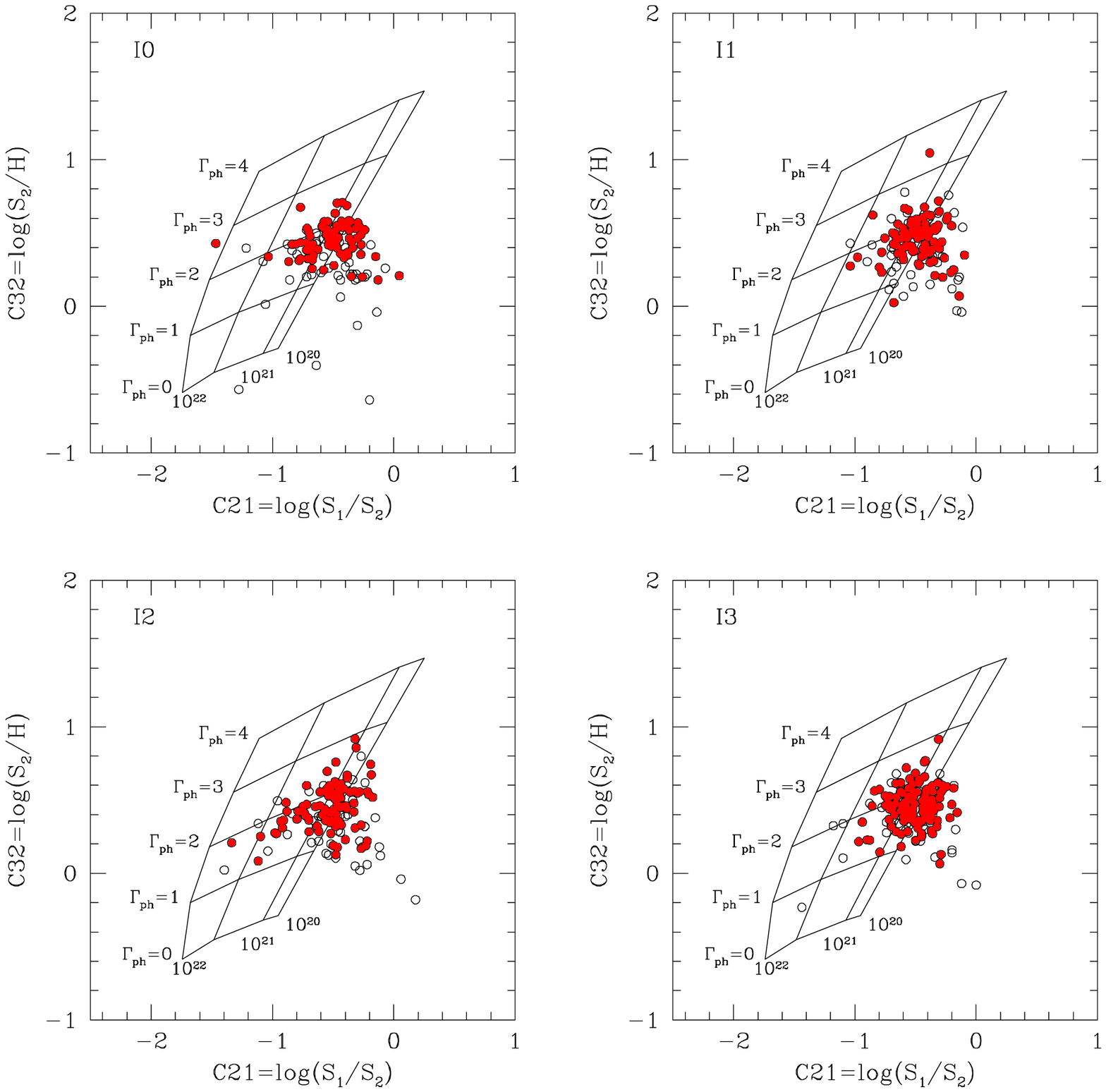}
\figcaption[f20.eps]{
The color-color diagram of ChaMP X-ray point sources observed with
ACIS-I. The open circles and red closed circles represent sources with 
$S/N>1.5$ and  with $S/N>2.0$, respectively.
The grid indicates the predicted
locations of the sources at redshift $z=0$ with various photon indices
($0 \leq \Gamma_{ph} \leq 4$, from bottom to top) and absorption column densities
($10^{20}\leq N_{H} \leq 10^{22}~cm^{-2}$, from right to left).
Most sources are located within the ranges of Galactic absorption  
$10^{20} \lesssim N_{H} \lesssim 10^{21}~cm^{-2} $ 
and photon index $1\lesssim\Gamma_{ph}\lesssim2.5$.
\label{fig-color-acisi}}
\clearpage

\plotone{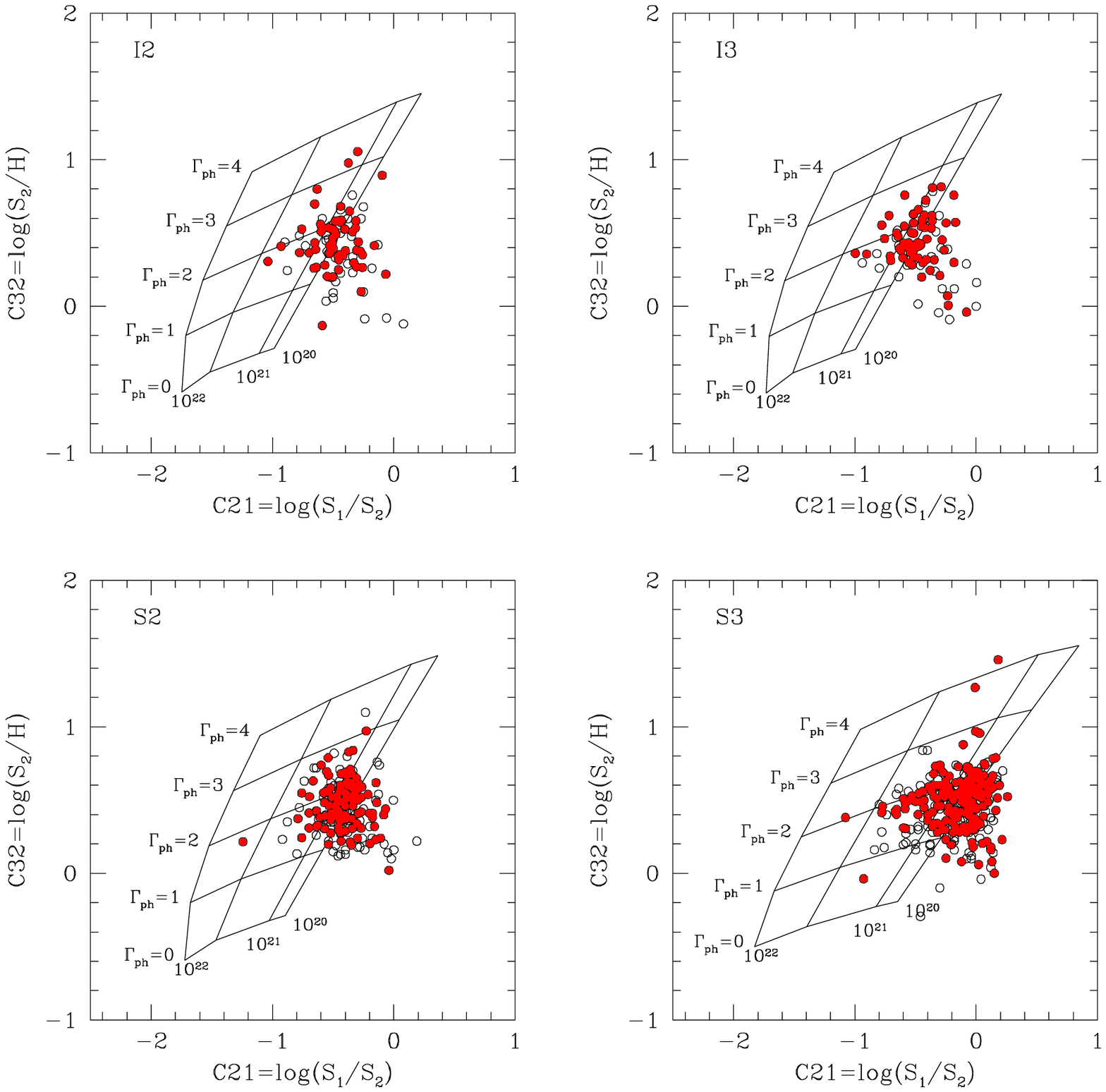}
\figcaption[f21.eps]{
Same as Figure \ref{fig-color-acisi}, but for ACIS-S observations.
\label{fig-color-aciss}}
\clearpage

\end{document}